\documentclass[12pt]{article}
\usepackage[cp1250]{inputenc}
\usepackage{amsmath}
\usepackage{amssymb}
\usepackage{graphicx}

\newtheorem{Theorem}{Theorem}[section]

\newtheorem{Lemma}[Theorem]{Lemma}

\newenvironment{Proof}[1]{{\bf Proof #1.} }{$\Box$\\}

\numberwithin{equation}{section}

\begin{document}

\centerline{\Large Thorough analysis of the Oseen system}

\smallskip

\centerline{\Large in 2D exterior domains.}

\smallskip

\bigskip

\smallskip

\centerline{Pawe\l ~ Konieczny}

\begin{center}

{Institute of Applied Mathematics and Mechanics}

{Warsaw University}

{ul. Banacha 2, 02-097 Warszawa, Poland}

{E-mail: konieczny@hydra.mimuw.edu.pl}

\end{center}

{\bf Abstract.} 
We construct $L_p$-estimates for the inhomogeneous Oseen system
studied in a two dimensional exterior domain $\Omega$ with
inhomogeneous slip boundary conditions.
The kernel of the paper is a result for the half space $\mathbb{R}^2_+$. Analysis of this model
system shows us a parabolic character of the studied problem, resulting as an appearance 
of the wake region behind the obstacle.
Main tools are given by the Fourier analysis to obtain the maximal regularity estimates.
The results imply the solvability for the Navier-Stokes system
for small velocity at infinity.
\medskip



\section{Introduction}


One of the main problems in the theory of the Navier-Stokes equations studied in exterior
domains is the question about the behaviour of the velocity vector field of the fluid at infinity.
The typical system in a bounded domain:
\begin{eqnarray}
 v\cdot\nabla v -\Delta v + \nabla p & = & F \qquad \textrm{in~} \Omega,\label{intr_NS_0}\\
 \textrm{div~} v & = & 0 \qquad \textrm{in~} \Omega,\label{intr_NS_10}\\
  B(v, p) & = & b	\qquad \textrm{on~} \partial\Omega,\label{intr_NS_20}
\end{eqnarray}
where $B(v, p)$ stands for the boundary constraints (e.g. Dirichlet boundary condition),
is complemented with the condition on the velocity vector field at infinity, namely
\begin{equation}\label{intr_NS_30}
	v\to \vec{v}_\infty \qquad\textrm{as~} |x|\to\infty
\end{equation}
for some prescribed constant vector field $v_\infty$. There are classical results of Leray
about existence of solutions with the finite Dirichlet integral to the system (\ref{intr_NS_0})-(\ref{intr_NS_20}). However
one cannot predict that these solutions satisfy (\ref{intr_NS_30}). Indeed, in two dimensions
we cannot use standard embedding theorems, since the dimension of the domain coincides with the power $2$
in the integral, that is why the condition 
\begin{equation}\label{Dir}
	\int_{\Omega} |\nabla v|^2 dx < \infty
\end{equation}
itself is insufficient even to assure that $v \in L^\infty(\Omega)$. The condition (\ref{Dir})
implies that $v\in BMO(\Omega)$ only, hence we are not able to deduce information
about the behaviour of $v$ at infinity.

Many mathematicians were investigating the problem (\ref{intr_NS_0})-(\ref{intr_NS_30})
and some partial results were obtained for example by Gilbarg and Weinberger (\cite{GW1}, \cite{GW2})
and Amick (\cite{Amick1988}). More general result was obtained by Finn and Smith (\cite{FiSm}) and
Galdi (\cite{Galdi1993}), where the key tool to assure that
(\ref{intr_NS_30}) holds, was a proper
$L_p$-estimate for the Oseen system considered with Dirichlet boundary constrains and an argument of fixed
point theorem. For a detailed discussion of these results we refer the Reader to (\cite{Galdi}).

The Oseen system has this advantage over the Stokes system that one can
obtain better information of the solution $v$ at infinity, because of the presence
of the additional term $v_{,1}$ (see  \cite{FiSm}, \cite{Galdi1993}).

While existence of solutions to this problem itself is very interesting also investigating their behaviour,
both close to the obstacle and at infinity, brings up many substantial questions. For example
what is the decay rate for the velocity at infinity and if there exists a wake region
behind the obstacle. Both of these questions have answers, depending on a proper information
of the solution (see \cite{GS}). One can expect that the decay rate of the solution to the Navier-Stokes
system will be similar to the decay rate of the Oseen fundamental solution,
however we do not want to address this question in our paper.

Our analysis of the Oseen system shows that the behaviour
of the solution depends strongly on the angle between the surface and the vector $\vec{v}_\infty$.
In a simplified case of a convex obstacle it can be shown, that the character of the system
is elliptic in front of the obstacle, while its character changes into parabolic type
behind the obstacle. This is presented on the following figure:
\begin{center}
	\includegraphics[width=9cm]{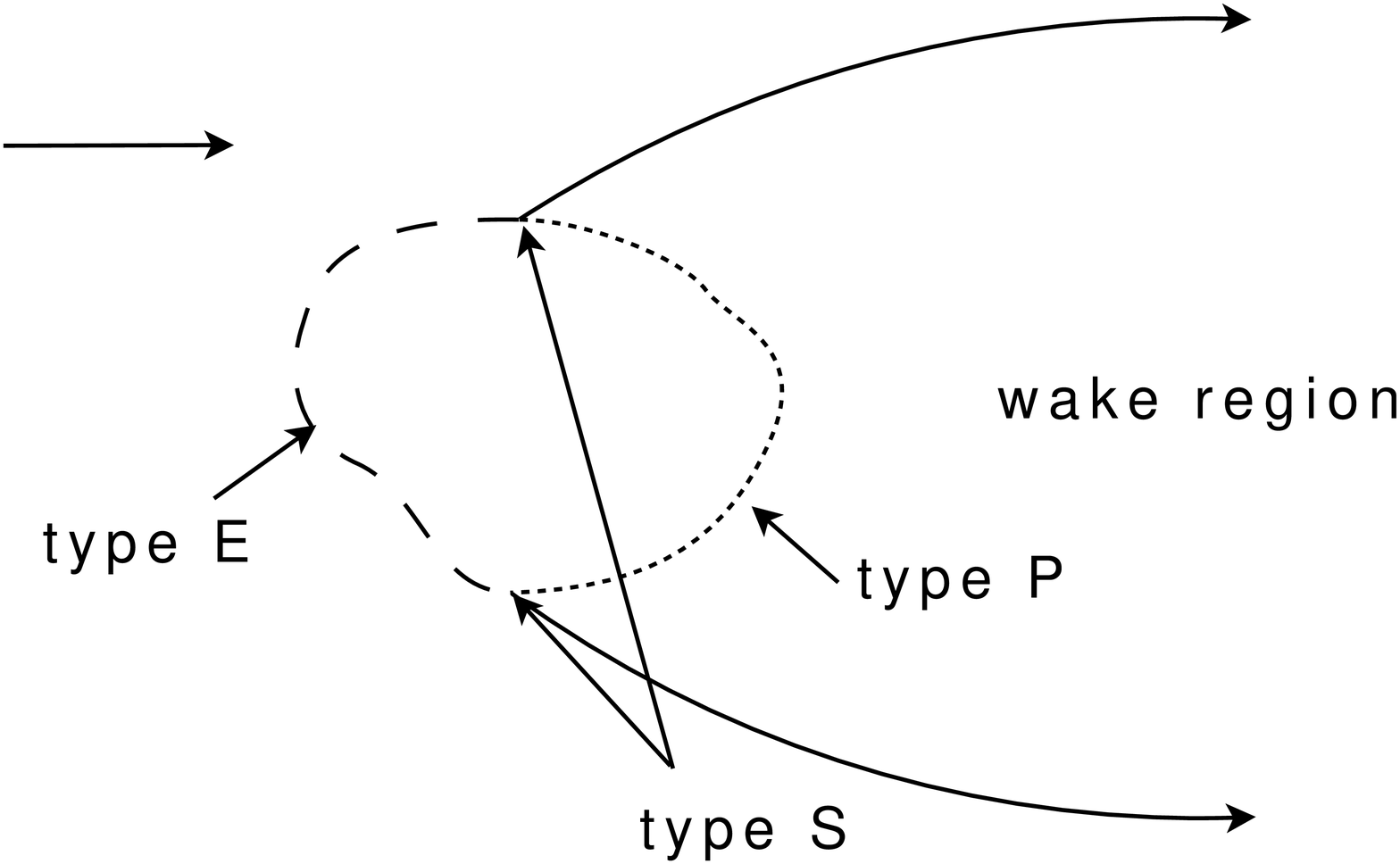}
\end{center}

The core  of the paper is the thorough analysis of the Oseen system in the half plane. 
We show that proper $L_p$--estimates are valid for the second derivatives of the velocity
and for the gradient of the pressure, under assumption, that boundary constraints
are in a suitable class of regularity. What is substantial is  the class of regularity
required by the boundary problem. 
It turns out that the choice of boundary data should depend on the sign of $\vec{v}_\infty \cdot\vec{n}$
($\vec{n}$ is the normal vector to the boundary),
which corresponds to the position of the obstacle.
In the case $\vec{v}_\infty \cdot\vec{n} < 0$ (points E), which analyzes the system in front
of the obstacle, the class of regularity of boundary data is of the elliptic type.
For $\vec{v}_\infty \cdot\vec{n} > 0$ (points P --- behind the obstacle) it appears
that the system loses its purely elliptic character in favour of a parabolic degeneration.
This feature corresponds to the appearance of the wake region behind the obstacle.
As $\vec{v}_\infty \cdot\vec{n} = 0$ (points S) we obtain a transition area.

As an application to this analysis we show $L_p$--estimates for the Oseen
system in exterior domain, which allows one to obtain also existence results for the Navier-Stokes system,
which by results of Galdi and Sohr \cite{GS} describes the structure of solutions at infinity.

We would like to emphasize that our aproach does not require explicit form of the fundamental solution. 
A similar approach has been examined by Solonnikov (\cite{Solonnikov}) and later by Zaj\c{a}czkowski and Mucha (\cite{ZaMuJDE}, \cite{ZaMuStud}).

Let us precise our problem. We consider the system:
\begin{eqnarray}
 v_\infty v_{,1} - \Delta v + \nabla p & = & F \qquad \textrm{in~} \Omega \label{intr0},\\
 {\textrm{div~}}v & = & G \qquad \textrm{in~} \Omega \label{intr10},\\
 \vec{n}\cdot\mathbb{T}(v, p)\cdot\vec{\tau}+f(v\cdot\vec{\tau}) & = & b \qquad \textrm{on~} \partial\Omega \label{intr20},\\
 v\cdot\vec{n} & = & d \qquad \textrm{on~} \partial\Omega \label{intr30},
\end{eqnarray}
together with a condition at infinity
\begin{equation}\label{intr40}
v\to \vec{v}_\infty \qquad \textrm{as~}|x|\to\infty,
\end{equation}
where the pair $(v, p)$ is the sought solution -- respectively the velocity vector field and the corresponding
pressure, $F$ is an external force acting on the fluid, $G$ is the function describing compressibility
of the fluid, $v_\infty$ is a constant describing the velocity of the fluid at inifinity,
$f$ is a nonnegative friction coefficient, $\mathbb{T}(v, p)$ is the Cauchy stress tensor, i.e.
$\mathbb{T}(v, p) = \nu\mathbb{D}(v) + p\mathbb{I}$, where
$\mathbb{D}(v) = \{v_{i, j} + v_{j, i}\}_{i, j = 1}^{2}$ is the symmetric
part of the gradient $\nabla v$, and $\mathbb{I}$ is the identity matrix.
Moreover $\vec{n}, \vec{\tau}$ are, respectively, the normal and tangential
vector to boundary $\partial\Omega$ of an exterior domain $\Omega$,
where $\Omega = \mathbb{R}^2\setminus B$, for a bounded
simply-connected domain $B\subset \mathbb{R}^2$. 

The slip boundary conditions govern the motion of particles at the boundary -- relation
(\ref{intr20}) is just Newton's second law. From the physical point
of view this constraint is more general than the Dirichlet boundary data, since
for $f\to\infty$ and $b\equiv 0$ one can obtain relation $v_{\partial\Omega } = 0$.
The case where $f = 0$ is important for applications, since then the fluid
reacts with surface $\partial\Omega $ as the perfect gas (\cite{Mucha1}).

In many modern applications, as the model of motion of blood, polymers and liquid
metals, this type of boundary conditions is widely used (\cite{Fujita}, \cite{Itoh}). 
Our considerations in an exterior 
domain are also important for example in the field of aerodynamics, where problems
with flow past an obstacle is of high interest. 

As a direct result of our analysis of the system in the halfplane we prove  the following theorem:
\begin{Theorem}\label{mainTh}
  Let $1 < p < \infty$, $F\in L^p(\Omega)\cap H^{-1}(\Omega)$, $G\in W^{1}_p(\Omega)$ 
  with $F$ and $G$ of compact support in $\Omega $, 
  $b\in W^{1-1/p}_p(\partial\Omega)$ 
  and $d\in W^{2-1/p}_p(\partial\Omega)$, 
  for which the following compatibility condition is fulfilled:
  \begin{equation}
  	\int_{\partial\Omega} d(x) d\sigma = \int_{\Omega} G dx.
  \end{equation}
  Moreover let $f > 0$ be a positive constant and $v_\infty \neq 0$. Then there exists a solution $(v, p)$ to the system
  (\ref{intr0})-(\ref{intr30}), for which the following estimate holds:
  \begin{equation}\label{intr50}
  	\begin{array}{rcl}
  		v_\infty\|v_{,1}\|_{L^p(\Omega)} & + & \|\nabla^2 v\|_{L^p(\Omega)}  +  \|\nabla p\|_{L^p(\Omega) } \leq  \\
  	  	& \leq & C(\Omega, f, v_\infty)\left(\|F\|_{L^p(\Omega)\cap H^{-1}(\Omega)} + \|G\|_{W^{1}_p(\Omega)} +\right. \\
  		& + & \left. \|b\|_{W^{1-1/p}_p(\Omega)} + \|d\|_{W^{2-1/p}_p(\Omega)}\right).
  	 \end{array}
  \end{equation}
  Denoting the term on the right hand side of (\ref{intr50}) as $C(DATA)$ we also have:
  \begin{itemize}
	\item for  $1 < p < 3$:	$v_\infty^{1/3} \|\nabla v\|_{L^{3p/(3-p)}} \leq C(\textrm{DATA})$
	\item for  $1 < p < 3/2$: $v_\infty^{2/3}\|v\|_{L^{3p/(3-2p)}} \leq C(\textrm{DATA})$.
\end{itemize}
\end{Theorem}

As was mentioned before, we may use this result together with the techniques from the work
of Galdi (\cite{Galdi1993}) to obtain the following result:
\begin{Theorem}\label{NSTh}
	Considering the system (\ref{intr_NS_0})-(\ref{intr_NS_10}) of Navier-Stokes equations in an exterior domain $\Omega $,
	together with slip boundary conditions (\ref{intr20})-(\ref{intr30}).
	where $b\in W^{1-1/p}_p(\partial\Omega)$ and $d\in W^{2-1/p}_p(\partial\Omega)$ and $v_\infty \neq 0$.
	If $d(x)$ satisfies the compatibility condition $\int_{\partial\Omega } d(x) d\sigma = 0$,
	then for $1 < p < 6/5$ and sufficiently small data (i.e. $\|b\|_{W^{1-1/p}_p(\partial\Omega )}$,
	$\|d\|_{W^{2-1/p}_p(\partial\Omega )}$, $v_\infty$) there exists a unique solution $(v, p)$ such 
	that:
	\begin{equation}
		v-v_\infty \in L^{3p/(3-2p)}(\Omega ),\quad
		\nabla v\in L^{3p/(3-p)}(\Omega),\quad
		\nabla^2 v\in L^p(\Omega ),
	\end{equation}
	and suitable estimates hold.
\end{Theorem}
We refer the Reader to \cite{Galdi} for a detailed discussion of this and similar results (for
example uniqueness of solution $(v, p)$ under suitable condition).

\textbf{Notation.}
Throughout the paper we use standard notation (\cite{Triebel}): $W^k_p(\Omega)$ for Sobolev spaces 
and the following definition of the norm in Slobodeckii spaces $W^s_p(\mathbb{R}^n)$:
\begin{multline}\label{not0}
	\|f\|_{W^s_p(\mathbb{R}^n)}^p =
		\int_{\mathbb{R}^n}|f(x)|^p dx +
		\sum_{0\leq|s'|\leq [|s|]} \int_{\mathbb{R}^n} |D^{s'} f(x)|^p dx+ 
		\|f\|_{\dot W^s_p(\mathbb{R}^n)}^p,
\end{multline}
where $[\alpha]$ stands for the integral part of $\alpha$ and
\begin{equation}\label{not10}
	\|f\|_{\dot W^s_p(\mathbb{R}^n)}^p = \sum_{|s'|=[|s|]} \int\limits_{\mathbb{R}^n\times\mathbb{R}^n}
		\frac{|D^{s'}f(x) - D^{s'}f(x')|^p}{|x-x'|^{n+p(|s|-[|s|])}}.
\end{equation}
We also introduce the following notation for intersected spaces. Let $X_r(U)$ be a Banach
space, dependent of a constant $r$, equipped with the norm $\|\cdot\|_{X_r(U)}$. Let $A \subset \mathbb{R}$
be a nonempty set. Then we introduce the following function space:
\begin{equation}\label{not30}
	X_{r, A}(U) = \bigcap_{r\in A} X_r(U),
\end{equation}
equipped with the norm
\begin{equation}\label{not40}
	\|f\|_{X_{r, A}(U)} = \sup_{r\in A} \|f\|_{X_r(U)}.
\end{equation}
In our case the set $A$ will be always of finite elements.

The structure of the paper is as follows: the core part, Section \ref{OseenR2p}, is devoted to the case 
of a flow in the halfplane. At the beginning we give some preliminary considerations,
we state main results in Theorems  \ref{R2pTheorem} and \ref{R2pTheoremV}
and then we give some results about consistency of the boundary conditions.
Later we derive a solution to our problem and give estimates for the pressure,
i.e. we prove Theorem \ref{R2pTheorem}. Section \ref{secondD} consists of 
introducing auxiliary problem for the velocity of the fluid. This result
is used to prove estimates for the second derivatives of $u$ in 
Section \ref{estU11} and Section \ref{estU22}, i.e. $u_{,11}$ and $u_{,22}$
respectively, which proves Theorem \ref{R2pTheoremV}.
At the end of Section \ref{OseenR2p} we give a brief summary
about the choice of boundary conditions.

Section \ref{sect:OR2} is devoted to results in the whole space $\mathbb{R}^2$,
which were being used in the previous section. 
As a consequence of results from Section \ref{OseenR2p} in part 4 we give a proof of Thorem \ref{mainTh}.
In Appendix we present two multiplier theorems of Marcinkiewicz type. We also
give some additional results, which are needful for our considerations, but are 
connected with a general theory of function spaces rather than with a theory of fluid dynamics.


\section{The Oseen system in the halfspace $\mathbb{R}^2_+$}\label{OseenR2p}

The localization procedure obviously changes not only the domain our problem is considered in, but
also affects its structure. The substantial difference is that the term $v_{\infty}v_{,1}$ from (\ref{intr0})
transforms into $a_1 v_{,1} + a_2 v_{,2} = (a_1, a_2)\cdot\nabla v$, where $a_1^2 + a_2^2 = v_\infty^2 > 0$.
We emphasize this because the sign of $a_2$ will be crucial in our considerations, since it is
the same as a sign of $\vec{v}_\infty\cdot\vec{n}$, which, for a convex obstacle, reflects
the region of the considered situation, namely the case $a_2 < 0$ corresponds to a region of the boundary in front of the obstacle, 
while $a_2 > 0$ stands for the situation behind the obstacle.

In this section we consider the following system:
\begin{eqnarray}\label{OseenAB_NH}
  a_1 v_{,1} + a_2 v_{,2} - \Delta v + \nabla q & = & F \qquad \qquad \textrm{in~} \mathbb{R}^2_+,\label{lpa0}\\
  \nabla\cdot v & = & G\qquad \qquad \textrm{in~} \mathbb{R}^2_+,\label{lpa10}\\
  \vec{n}\cdot\mathbb{T}(v, p)\cdot\vec{\tau} + f(v\cdot\tau) & = & \underline{b}\qquad \qquad \textrm{on~} \partial\mathbb{R}^2_+,\label{lpa20}\\
  \vec{n}\cdot v & = & \underline{d} \qquad \qquad \textrm{on~} \partial\mathbb{R}^2_+,\label{lpa25} \\
  v & \to & 0\qquad \qquad \textrm{as~} |x|\to\infty.\label{lpa30}
 \end{eqnarray}
We assume that $F$ and $G$ have compact support in $\mathbb{R}^2_+$, since this system comes from the localization procedure.

To simplify the problem we remove the innhomogeneity from (\ref{lpa0}) and (\ref{lpa10}) using results
in the whole space $\mathbb{R}^2$, i.e. Theorem \ref{OR2_40}. Then we need to use Lemma \ref{OR2_75}
to see, in which class of regularity on the boundary of $\mathbb{R}^2_+$ the obtained solution is.
We gather this in the following Lemma:
\begin{Lemma}\label{extLemma}
	Let $q > 3$. Given $F \in L^q(\mathbb{R}^2)\cap L^1(\mathbb{R}^2)$ and $G \in W^1_q(\mathbb{R}^2)\cap W^1_1(\mathbb{R}^2)$.
	Considering	the following Oseen system in the whole space:
	\begin{eqnarray}
		a_1 \widetilde{v}_{,1} + a_2 \widetilde{v}_{,2} - \Delta \widetilde{v} + \nabla \widetilde{q} & = & \widetilde{F}
		\qquad \textrm{in~} \mathbb{R}^2,\label{OseenR2_1}\\
		\mathrm{div~}\widetilde{v} & = & \widetilde{G} \qquad \textrm{in~} \mathbb{R}^2,\label{OseenR2_2}
	\end{eqnarray}
	there exists a solution $(\widetilde{v}, \widetilde{q})$ for this system, for which the following
	conditions are satisfied:
	\begin{eqnarray}
		\widetilde{v} & \in & W^{2}_r(\mathbb{R}^2) \qquad \textrm{for all~}r\in (3, q],\label{boundext_20}\\
		\nabla\widetilde{v} & \in & W^1_r(\mathbb{R}^2) \qquad \textrm{for all~} r\in (3/2, q],\label{boundext_30}
	\end{eqnarray}
	and for all $r\in (3, q]$:
	\begin{eqnarray}
		 \vec{n}\cdot\mathbb{T}(\widetilde{v}, \widetilde{q})\cdot\vec{\tau}_{|x_2 = 0} + f(\widetilde{v}\cdot\vec{\tau})_{|x_2 = 0} & 
		 		\in & W^{1-1/r}_r(\mathbb{R}),\label{boundext_40}\\
		 \vec{n}\cdot \widetilde{v}_{|x_2 = 0} & \in & W^{2-1/r}_r(\mathbb{R}),\label{boundext_50}
	\end{eqnarray}
	like also for all $r\in (3/2, q]$:
	\begin{eqnarray}
		 \vec{n}\cdot\mathbb{T}(\widetilde{v}, \widetilde{q})\cdot\vec{\tau}_{|x_2 = 0} + f(\widetilde{v}\cdot\vec{\tau})_{|x_2 = 0} & 
		 		\in & \dot W^{1-1/r}_r(\mathbb{R}),\label{boundext_60}\\
		 \vec{n}\cdot \widetilde{v}_{|x_2 = 0} & \in & \dot W^{2-1/r}_r(\mathbb{R})\cap\dot W^{1-1/r}_r(\mathbb{R}).\label{boundext_70}
	\end{eqnarray}
\end{Lemma}

\begin{Proof}{}
	Existence of a solution $(\widetilde{v}, \widetilde{q})$ is straightforward from Theorem \ref{R2pTheorem}.
	Conditions (\ref{boundext_20})-(\ref{boundext_30}) come from (\ref{OR2_76}) and (\ref{OR2_77}).
	(\ref{boundext_20}) immediately imply (\ref{boundext_40})-(\ref{boundext_50}). Condition
	(\ref{boundext_30}) implies that $\vec{n}\cdot\mathbb{T}(\widetilde{v}, \widetilde{q})\cdot\vec{\tau} \in \dot W^{1-1/r}_r(\mathbb{R})$,
	however to show $f(\widetilde{v}\cdot\vec{\tau})\in \dot W^{1-1/r}_r(\mathbb{R})$ one has
	to use (\ref{OR2_78}). Condition (\ref{boundext_70}) also comes from (\ref{OR2_78}).	
\end{Proof}

Using the above Lemma we are able to simplify the system (\ref{lp0})-(\ref{lp30}). Denoting $v = u + \widetilde{v}$ and $q = p + \widetilde{q}$ we get a system for $(u, p)$
\begin{eqnarray}\label{OseenAB}
  a_1 u_{,1} + a_2 u_{,2} - \Delta u + \nabla p & = & 0 \qquad \qquad \textrm{in~} \mathbb{R}^2_+,\label{lp0}\\
  \nabla\cdot u & = & 0\qquad \qquad \textrm{in~} \mathbb{R}^2_+,\label{lp10}\\
  \vec{n}\cdot\mathbb{T}(u, p)\cdot\vec{\tau} + f(v\cdot\tau) & = & b\qquad \qquad \textrm{on~} \partial\mathbb{R}^2_+,\label{lp20}\\
  \vec{n}\cdot u & = & d \qquad \qquad \textrm{on~} \partial\mathbb{R}^2_+,\label{lp25}\\
  u &\to & 0\qquad \qquad \textrm{as~} |x|\to\infty.\label{lp30}
\end{eqnarray}
where for the readability we denoted the term 
$(\underline{b} - \vec{n}\cdot\mathbb{T}(\widetilde{v}, \widetilde{q})\cdot\vec{\tau} - f(\widetilde{v}\cdot\vec{\tau}))$ as $b$ and $\underline{d} - \vec{n}\cdot \widetilde{v}$ as $d$.

The main result concernes estimates for the pressure and for the velocity. For the readability
of the paper we split it into two theorems within each we consider some cases.
We emphasize that we use homogeneous spaces $\dot W^s_p$ and inhomogeneous spaces $W^s_p$.
\begin{Theorem}{\emph{\textbf{Estimates for the pressure.}}}\label{R2pTheorem}
  Let $f > 0$ be a constant friction coefficient and $p > 3/2$. 
  Given the solution $(u, p)$ to the system (\ref{lp0})-(\ref{lp30}). Considering the following cases: 
  \begin{itemize}
  	\item for $a_2 \leq 0$: let $b \in \dot W^{1-1/r}_r(\mathbb{R})$ and $d\in \dot W^{2-1/r}_r(\mathbb{R})\cap \dot W^{1-1/r}_r(\mathbb{R})$ for all $3/2 < r \leq p$.
  		Then $\nabla p \in L^r(\mathbb{R}^2_+)$ for all $3/2 < r \leq p$ and the following inequality holds:
	  \begin{equation}
	  	\|\nabla p\|_{L^r(\mathbb{R}^2_+)} 
	  		\leq C(f, a_1, a_2)\left(\|b\|_{\dot W^{1-1/r}_r(\mathbb{R})} + 
	  					\|d\|_{\dot W^{2-1/r}_r(\mathbb{R})\cap \dot W^{1-1/r}_r(\mathbb{R})}\right).
	  \end{equation}
  	\item for $a_2 > 0$: let $b \in \dot W^{1-1/r}_r(\mathbb{R})$ and $d\in \dot W^{2-1/r}_r(\mathbb{R})$ for all $3/2 < r \leq p$.
  		Then $\nabla p \in L^r(\mathbb{R}^2_+)$ for all $3/2 < r \leq p$ and the following inequality holds:
	  \begin{equation}
	  	\|\nabla p\|_{L^r(\mathbb{R}^2_+)} 
	  		\leq C(f, a_1, a_2)\left(\|b\|_{\dot W^{1-1/r}_r(\mathbb{R})} + 
	  					\|d\|_{\dot W^{2-1/r}_r(\mathbb{R})}\right).
	  \end{equation}
  \end{itemize}
\end{Theorem}
\begin{Theorem}{\emph{\textbf{Estimates for the velocity.}}}\label{R2pTheoremV}
  Let $f > 0$ be a constant friction coefficient and $p > 3$.
  Given the solution $(u, p)$ to the system (\ref{lp0})-(\ref{lp30}). Considering the following cases:
  \begin{itemize}
  	\item for $a_2 < 0$: let $b \in W^{1-1/r}_r(\mathbb{R})$ and $d\in W^{2-1/r}_r(\mathbb{R})$ for all $3/2 < r \leq p$. Then
  		$\nabla^2 u \in L^s(\mathbb{R}^2_+)$ for all $s\in (3, p]$ and the following inequality holds:
  		\begin{equation}
  		 	\|\nabla^2 u\|_{L^s(\mathbb{R}^2_+)}  \leq C(f, a_1, a_2)\left(\|b\|_{W^{1-1/s}_{s, A_s}(\mathbb{R})} + \|d\|_{W^{2-1/s}_{s, A_s}(\mathbb{R})}\right),
  		\end{equation} 
  		where $A_s = \{3s/(3+s), s\}$
  	\item for $a_2 = 0$: let $b \in \dot W^{1-1/r}_r(\mathbb{R})$ and $d\in \dot W^{2-1/r}_r(\mathbb{R})\cap \dot W^{1-1/r}_r(\mathbb{R})$ for all $3/2 < r \leq p$. Then
  		$\nabla^2 u \in L^s(\mathbb{R}^2_+)$ for all $s\in (3, q]$ and the following inequality holds:
  		\begin{equation}
  		 	\|\nabla^2 u\|_{L^s(\mathbb{R}^2_+)}  \leq C(f, a_1, a_2)\left(\|b\|_{\dot W^{1-1/s}_{s, A_s}(\mathbb{R})} + \|d\|_{\dot W^{2-1/s}_{s, A_s}(\mathbb{R})\cap \dot W^{1-1/s}_{s, A_s}(\mathbb{R})}\right),
  		\end{equation} 
  		where $A_s = \{3s/(3+s), s\}$.
  	\item for $a_2 > 0$: let $b \in \dot W^{1-1/r}_r(\mathbb{R}) \cap \dot W^{1-2/r}_r(\mathbb{R})$ and 
  		$d\in \dot W^{2-1/r}_r(\mathbb{R})\cap \dot W^{2-2/r}_r(\mathbb{R})$ for all $2 < r \leq p$. Then
  		$\nabla^2 u \in L^s(\mathbb{R}^2_+)$ for all $s\in (3, q]$ and the following inequality holds:
  		\begin{equation}
  		 	\|\nabla^2 u\|_{L^s(\mathbb{R}^2_+)}  \leq C(f, a_1, a_2)\left(\|b\|_{\dot W^{1-1/s}_{s, A_s}(\mathbb{R})} + \|d\|_{\dot W^{2-1/s}_{s, A_s}(\mathbb{R})\cap \dot W^{2-1/s}_{s, A_s}(\mathbb{R})}\right),
  		\end{equation} 
  		where $A_s = \{3s/(3+s), s\}$.
  \end{itemize}
  
\end{Theorem}

\textbf{Remark:} We would like to emphasize, that 
above assumptions on $b$ and $d$ are consistent with Lemma \ref{extLemma} in the sense, that
the procedure of subtracting the inhomogeneity from the right hand side of (\ref{lpa0})-(\ref{lpa10})
does not determine the regularity of boundary conditions $b$ and $d$. We discuss this in details: let
us recall that $b$ and $d$ from the right hand side of (\ref{lp20}) and (\ref{lp25}) come from
the subtraction of terms $\vec{n}\cdot\mathbb{T}(\widetilde{v}, \widetilde{q})\cdot\vec{\tau} + f(\widetilde{v}\cdot\vec{\tau}) $
and $\vec{n}\cdot \widetilde{v}$ from $\underline{b}$ and $\underline{d}$ from the original boundary constraints (\ref{lpa20}) and (\ref{lpa25}).
Let us thus denote $\vec{n}\cdot\mathbb{T}(\widetilde{v}, \widetilde{q})\cdot\vec{\tau} + f(\widetilde{v}\cdot\vec{\tau}) $
as $\widetilde{b}$ and $\vec{n}\cdot \widetilde{v}$ as $\widetilde{d}$ and check if assumptions
on boundary conditions in Theorem \ref{R2pTheorem} and Theorem \ref{R2pTheoremV} are consistent with
regularity of $\widetilde{b}$ and $\widetilde{d}$.

In Theorem \ref{R2pTheorem} we assume $b\in \dot W^{1-1/r}_r(\mathbb{R})$ and $d\in \dot W^{2-1/r}_r(\mathbb{R})$
for all $3/2 < r \leq p$ in case $a_2 > 0$, and $b\in \dot W^{1-1/r}_r(\mathbb{R})$ and 
$d\in \dot W^{2-1/r}_r(\mathbb{R})\cap \dot W^{1-1/r}_r(\mathbb{R})$ in case $a_2 \leq 0$, but due to
(\ref{boundext_60})-(\ref{boundext_70}) we have $\widetilde{b} \in \dot W^{1-1/r}_r(\mathbb{R})$
amd $\widetilde{d} \in \dot W^{2-1/r}_r(\mathbb{R})\cap \dot W^{1-1/r}_r(\mathbb{R})$
for all $r\in (3/2, p]$, and hence subtraction of $\widetilde{b}$ and $\widetilde{d}$
does not influence regularity of $b$ and $d$ and one has the following inequalities:
\begin{itemize}
	\item for $a_2 > 0$ and all $r \in (3/2, p]$:
		\begin{eqnarray}
			\|b\|_{\dot W^{1-1/r}_r(\mathbb{R})} & \leq & C(\|\underline{b}\|_{\dot W^{1-1/r}_r(\mathbb{R})} + 
						\|F\|_{L^r_{A_r}(\mathbb{R}^2_+)} + \|G\|_{W^1_{r, A_r}(\mathbb{R}^2_+)}),\\
			\|d\|_{\dot W^{2-1/r}_r(\mathbb{R})} & \leq & C(\|\underline{d}\|_{\dot W^{1-1/r}_r(\mathbb{R})} + 
						\|F\|_{L^r_{A_r}(\mathbb{R}^2_+)} + \|G\|_{W^1_{r, A_r}(\mathbb{R}^2_+)}),
		\end{eqnarray}
		where $A_r = \{3r/(3+r), r\}$,
	\item for $a_2 \leq 0$ and all $r \in (3/2, p]$:
		\begin{multline}
				\|b\|_{\dot W^{1-1/r}_r(\mathbb{R})} \leq C(\|\underline{b}\|_{\dot W^{1-1/r}_r(\mathbb{R})} + 
							\|F\|_{L^r_{A_r}(\mathbb{R}^2_+)} + \|G\|_{W^1_{r, A_r}(\mathbb{R}^2_+)}),		
		\end{multline}
		\begin{multline}
				\|d\|_{\dot W^{2-1/r}_r(\mathbb{R})\cap \dot W^{1-1/r}_r(\mathbb{R})}  \leq  
							C(\|\underline{d}\|_{\dot W^{2-1/r}_r(\mathbb{R})\cap \dot W^{1-1/r}_r(\mathbb{R})} + \\
							\|F\|_{L^r_{A_r}(\mathbb{R}^2_+)} + \|G\|_{W^1_{r, A_r}(\mathbb{R}^2_+)}),					
		\end{multline}
		where $A_r = \{3r/(3+r), r\}$.	
\end{itemize}

Similarly, using properties (\ref{boundext_40}) and (\ref{boundext_50}), we are able to show the following inequalities:
\begin{itemize}
	\item for $a_2 > 0$ and all $r \in (3, p]$:
		\begin{multline}
				\|b\|_{\dot W^{1-1/r}_r(\mathbb{R})\cap \dot W^{1-2/r}_r(\mathbb{R})}  \leq  
							C(\|\underline{b}\|_{\dot W^{1-1/r}_r(\mathbb{R})\cap \dot W^{1-2/r}_r(\mathbb{R})} + \\
							\|F\|_{L^r_{A_r}(\mathbb{R}^2_+)} + \|G\|_{W^1_{r, A_r}(\mathbb{R}^2_+)}),					
		\end{multline}
		\begin{multline}
				\|d\|_{\dot W^{2-1/r}_r(\mathbb{R})\cap \dot W^{2-2/r}_r(\mathbb{R})}  \leq  
							C(\|\underline{d}\|_{\dot W^{2-1/r}_r(\mathbb{R})\cap \dot W^{2-2/r}_r(\mathbb{R})} + \\
							\|F\|_{L^r_{A_r}(\mathbb{R}^2_+)} + \|G\|_{W^1_{r, A_r}(\mathbb{R}^2_+)}),					
		\end{multline}		
		where $A_r = \{3r/(3+2r), 3r/(3+r), r\}$,
	\item for $a_2 = 0$ and all $r \in (3, p]$:
		\begin{multline}
				\|b\|_{\dot W^{1-1/r}_r(\mathbb{R})}  \leq  
							C(\|\underline{b}\|_{\dot W^{1-1/r}_r(\mathbb{R})\cap \dot W^{1-2/r}_r(\mathbb{R})} + \\
							\|F\|_{L^r_{A_r}(\mathbb{R}^2_+)} + \|G\|_{W^1_{r, A_r}(\mathbb{R}^2_+)}),					
		\end{multline}
		\begin{multline}
				\|d\|_{\dot W^{2-1/r}_r(\mathbb{R})\cap \dot W^{1-1/r}_r(\mathbb{R})}  \leq  
							C(\|\underline{d}\|_{\dot W^{2-1/r}_r(\mathbb{R})\cap \dot W^{2-2/r}_r(\mathbb{R})} + \\
							\|F\|_{L^r_{A_r}(\mathbb{R}^2_+)} + \|G\|_{W^1_{r, A_r}(\mathbb{R}^2_+)}),					
		\end{multline}		
		where $A_r = \{3r/(3+r), r\}$,
	\item for $a_2 < 0$ and all $r \in (3, p]$:
		\begin{multline}
				\|b\|_{W^{1-1/r}_r(\mathbb{R})}  \leq  
							C(\|\underline{b}\|_{W^{1-1/r}_r(\mathbb{R})} + 
							\|F\|_{L^r_{A_r}(\mathbb{R}^2_+)} + \|G\|_{W^1_{r, A_r}(\mathbb{R}^2_+)}),					
		\end{multline}
		\begin{multline}
				\|d\|_{W^{2-1/r}_r(\mathbb{R})}  \leq  
							C(\|\underline{d}\|_{W^{2-1/r}_r(\mathbb{R})} + 
							\|F\|_{L^r_{A_r}(\mathbb{R}^2_+)} + \|G\|_{W^1_{r, A_r}(\mathbb{R}^2_+)}),					
		\end{multline}		
		where $A_r = \{3r/(3+2r), 3r/(3+r), r\}$.		
\end{itemize}

\subsection{Derivation of the solution.}

Regularity results from Theorem \ref{R2pTheorem} come from the formula for the solution, while
to show estimates from Theorem \ref{R2pTheoremV} we will consider auxiliary system. 

In this section we derive the solution using the Fourier transform
and solving algebraically the obtained system of ODEs. 

\noindent Let 
\begin{equation}\label{lp35}
v(\xi_1, x_2) = \mathcal{F}_{x_1}(u)\quad \textrm{and}\quad\pi(\xi_1, x_2) = \mathcal{F}_{x_1}(p),
\end{equation} 
where $\mathcal{F}_{x_1}$ is the Fourier transform
with respect to $x_1$, i.e.:
\begin{equation}\label{lp40}
  v(\xi_1, x_2) = \int_{\mathbb{R}} e^{-i\xi_1x_1}u(x_1, x_2) dx_1.
\end{equation}
First two equations of (\ref{OseenAB}) give us the following system:
\begin{eqnarray}
 a_1ik v_1 + a_2\dot v_{1} + k^2 v_1 - \ddot v_{1} + ik \pi & = & 0 \qquad \qquad \textrm{in~} \mathbb{R}^2_+,\label{lp80}\\
 a_1ik v_2 + a_2\dot v_{2} + k^2 v_2 - \ddot v_{2} + \dot\pi & = & 0 \qquad \qquad \textrm{in~} \mathbb{R}^2_+,\label{lp90}\\
 ik v_1 + \dot v_{2} & = & 0 \qquad \qquad \textrm{in~} \mathbb{R}^2_+.\label{lp100}
\end{eqnarray}
where we denoted $\partial_{x_2}$ as $\dot~$ ($x_2\to t$) and $\xi_1$ as $k$.

Solving this system we are interested in eigenvalues with negative real part. We thus have:
\begin{eqnarray}
 \lambda_1 & = & -|k|, \label{lp130}\\
 \lambda_3 & = & \frac{1}{2}(a_2-\sqrt{a_2^2+4(k^2+ia_1k)}),\label{lp150}
\end{eqnarray}
and we can present the solution to (\ref{lp80})-(\ref{lp100}) as
\begin{eqnarray}\label{}
 v_1(t, k) & = &  e^{t\lambda_1}U_{01}(k) + e^{t\lambda_3}U_{03}(k), \label{lp180}\\
 v_2(t, k) & = & -\frac{ik}{\lambda_1(k)}e^{t\lambda_1}U_{01}(k) - \frac{ik}{\lambda_3(k)}e^{t\lambda_3}U_{03},\label{lp190}\\
 \pi(t, k) & = & -(a_1+\sigma(k)ia_2)e^{t\lambda_1}U_{01}(k),\label{lp200}
\end{eqnarray}
where function $U_{0i}(k)$ are calculated from the boundary conditions $(\ref{OseenAB})_{3, 4}$ and are given
by the following formula:
\begin{eqnarray}
U_{01}(k) & = & \frac{\lambda_1(\hat{d}(k)(\lambda_3(-f+\lambda_3)+k^2)+ik \hat{b}(k))}{ik(\lambda_3-\lambda_1)(-f+\lambda_3+\lambda_1)},\label{lp250}\\
U_{03}(k) & = & \frac{\hat{b}(k) - ik\hat{d}(k) - (f-\lambda_1) U_{01}(k)}{(f-\lambda_3)}.\label{lp260}
\end{eqnarray}

Immediately, having this solution, we formulate the following result, which
gives us the reason to consider only Dirichlet boundary conditions:
\begin{Lemma}\label{boundLemma}
	Given a vector field $u$ such that  $v = \mathcal{F}_{x_1}(u)$
	satisfies (\ref{lp180})-(\ref{lp190}) for some functions $U_{01}(\xi)$ and $U_{03}(\xi)$.
	If $u$ satisfies the following slip boundary conditions
	\begin{equation}\label{lp262}
	  \begin{array}{rcl}
		\vec{n}\cdot\mathbb{D}(v)\cdot\vec{\tau} + f(v\cdot\vec{\tau}) & = & b,\\
		\vec{n}\cdot v & = & d,	    
	  \end{array}
	\end{equation}
	where $b\in \dot W^{1-1/p}_p(\mathbb{R})$, $d\in \dot W^{2-1/p}_p(\mathbb{R})$, 
	then this vector field satisfies also Dirichlet boundary constraints on $\partial \mathbb{R}^2_+$:
	\begin{equation}
		u(x_1, 0) = D(x_1),
	\end{equation}
	where $D\in \dot W^{2-1/p}_p(\mathbb{R})$ and is given by:
	\begin{equation}\label{boundLemma8}
	     \begin{array}{rcl}
			D_1(x_1) & = &  \mathcal{F}^{-1}_{\xi}\left(\frac{\hat{b}(\xi) - \hat{d}(\xi)(i\sigma(\xi)\lambda_3(\xi) - i\xi)}{f-\lambda_1(\xi) -\lambda_3(\xi)}\right)(x_1) \\
			D_2(x_1) & = & -d(x_1).
	     \end{array}
	\end{equation}
	and satisfies the following inequality:
	\begin{equation}
		\|D\|_{\dot W^{2-1/p}_p(\mathbb{R})} \leq C(\|b\|_{\dot W^{1-1/p}_p(\mathbb{R})} + \|d\|_{\dot W^{2-1/p}_p(\mathbb{R})}).
	\end{equation}
	Moreover, for $b\in \dot W^{1-1/p}_p(\mathbb{R})$ and $d\in \dot W^{2-1/p}_p(\mathbb{R})\cap \dot W^{1-1/p}_p(\mathbb{R})$ one has:
	\begin{equation}\label{boundLemma10}
		D\in \dot W^{2-1/p}_p(\mathbb{R})\cap \dot W^{1-1/p}_p(\mathbb{R}),
	\end{equation}
	and the following inequality holds:
	\begin{equation}
		\|D\|_{\dot W^{2-1/p}_p(\mathbb{R})\cap \dot W^{1-1/p}_p(\mathbb{R})} 
			\leq 
				C(\|b\|_{\dot W^{1-1/p}_p(\mathbb{R})} + \|d\|_{\dot W^{2-1/p}_p(\mathbb{R})\cap \dot W^{1-1/p}_p(\mathbb{R})}).
	\end{equation}

	\textbf{Remark:} Above homogeneous spaces can be replaced with inhomogeneous ones without any additional assumptions. 
\end{Lemma}

\begin{Proof}{of Lemma \ref{boundLemma}}
	The proof of this lemma is rather simple. While having exact formula (\ref{boundLemma8}) for $D_1$ and $D_2$
	one may use the Marcinkiewicz Theorem \ref{Marcinkiewicz} to obtain desired estimates.
\end{Proof}

\subsection{Estimate of the pressure.}

In this part of the paper we give a proof of Theorem \ref{R2pTheorem}.
Let us start now with estimates for the pressure
\begin{equation}
	p(x_1, x_2) = \mathcal{F}^{-1}_{k}\left(\pi(k, x_2)\right).\label{lp270}
\end{equation}
Recall (\ref{lp200}):
\begin{equation}
	\pi(t, k) = -(a_1+\sigma(k)ia_2)e^{t\lambda_1}U_{01}(k).
\end{equation}
We want to estimate the gradient $\nabla p$, i.e. $\partial_{x_1} p$ and $\partial_{x_2} p$.
These two terms correspond, after the Fourier transform, to terms $ik\pi(t, k)$ and $-|k|\pi(t, k)$,
which, from the point of view of our approach, are equivalent, since they differ
only by a function $\sigma(k)$, which, as we shall see, makes no difference in our estimates.

Let us thus focus on the term
\begin{equation}\label{lp275}
	ik\pi(t, k) = -e^{t\lambda_1}\frac{ik(a_1+\sigma(k)ia_2)\lambda_1(\hat{d}(k)(\lambda_3(-f+\lambda_3)+k^2)+ik \hat{b}(k))}{ik(\lambda_3-\lambda_1)(-f+\lambda_3+\lambda_1)}.
\end{equation}
We remove $(a_1+\sigma(k)ia_2)$ noticing that :
\begin{equation}
	\frac{ik}{\lambda_3-\lambda_1} = -\frac{a_2+2|k|+\sqrt{a_2^2+4(k^2+a_1ik)}}{a_1+\sigma(k)a_2i}.
\end{equation}
Thus we  present $ik\pi(t, k)$ as follows:
\begin{multline}\label{lp280}
	ik\pi(t, k) = \\
	= e^{t\lambda_1}\frac{(a_2+2|k|+\sqrt{a_2^2+4(k^2+a_1ik)})\lambda_1(\hat{d}(k)(\lambda_3(-f+\lambda_3)+k^2)+ik \hat{b}(k))}{ik(-f+\lambda_3+\lambda_1)}.
\end{multline}
First we focus on the term involving $\hat{b}(k)$.
Let
\begin{equation}\label{lp290}
	I_1(t, k) = e^{t\lambda_1}ik \frac{(a_2+2|k|+\sqrt{a_2^2+4(k^2+a_1ik)})\lambda_1}{ik(-f+\lambda_3+\lambda_1)}\hat{b}(k).
\end{equation}
and we denote as $I_2$ the remaining part of $ik\pi(t, k)$:
\begin{equation}\label{lp295}
	I_2(t, k) = e^{t\lambda_1}ik \frac{(a_2+2|k|+\sqrt{a_2^2+4(k^2+a_1ik)})\sigma(k)(\lambda_3(-f+\lambda_3)+k^2)}{ik(-f+\lambda_3+\lambda_1)}\hat{d}(k).
\end{equation}
We  present $I_1(t, k)$ as
\begin{equation}\label{lp300}
	I_1(t, k) = e^{t\lambda_1}ik \varphi_1(k) \hat{b}(k),
\end{equation}
where
\begin{equation}\label{lp310}
	\varphi_1(k) = \frac{(a_2+2|k|+\sqrt{a_2^2+4(k^2+a_1ik)})\lambda_1}{ik(-f+\lambda_3+\lambda_1)}.
\end{equation}
Since $\Re(\lambda_3+\lambda_1) \leq 0$ and $f > 0$ we have $\Re(-f + \lambda_3 + \lambda_1) \leq -f < 0$ and a
function $\varphi_1(k)$ is a proper multiplier in the sense of Theorem \ref{Marcinkiewicz} -- indeed, since
$\lambda_1 = -|k|$ our multiplier is bounded and smooth for $k\in \mathbb{R}\setminus \{0\}$. 
Moreover its derivative has a good decay rate, which guarantees that $|k|\cdot\varphi_{1}'(k)$
is bounded for all $k\in \mathbb{R}$.

The above considerations justify the following inequality:
\begin{equation}\label{lp360}
	\|\mathcal{F}^{-1}_{k}\left(I_{1}(t, \cdot)\right)\|_{L^p(\mathbb{R})} \leq 
		C\|\mathcal{F}^{-1}_{k}\left(e^{t\lambda_1} ik \hat{b}(k)\right)\|_{L^p(\mathbb{R})}.
\end{equation}
We now estimate the term:
\begin{equation}\label{lp370}
	\mathcal{F}^{-1}_{k}\left(-e^{t\lambda_1}ik \hat{b}(k)\right) =
	\mathcal{F}^{-1}_{k}\left(-e^{t\lambda_1}ik\right)\ast \mathcal{F}^{-1}_{k}\left(\hat{b}(k)\right),
\end{equation}
where $\ast$- is a convolution with respect to $x_1$.
Now since
\begin{equation}\label{lp380}
	\mathcal{F}^{-1}_{k}\left(-ike^{-|k|t}\right) = \left(\sqrt{\frac{2}{\pi}}\frac{t}{t^2+x_1^2}\right)_{,x_1} = -\sqrt{\frac{2}{\pi}}\frac{2x_1t}{(t^2+x_1^2)^2},
\end{equation}
we  rewrite our term as follows:
\begin{equation}\label{lp390}
	\mathcal{F}^{-1}_{k}\left(-e^{t\lambda_1}ik\right)\ast \mathcal{F}^{-1}_{k}\left(\hat{b}(k)\right) =
	\sqrt{\frac{2}{\pi}}\frac{x_1t}{(t^2+x_1^2)^2}\ast b(x_1) =: \sqrt{\frac{2}{\pi}}J_1(t, x_1).
\end{equation}

\noindent Now since
\begin{equation}\label{lp400}
	\int_{\mathbb{R}}b(x) \frac{-2yt}{(t^2+y^2)^2} dy = 0,
\end{equation}
we write:
\begin{equation}\label{lp410}
	J_1(t, x) = \int_{\mathbb{R}}\frac{2yt}{(t^2+y^2)^2}\left[b(x-y)-b(x)\right]dy.
\end{equation}
First we focus on:
\begin{equation}\label{lp420}
	\|J_1(t, \cdot)\|_{L^p(\mathbb{R})}^p = \left(
	\int_{\mathbb{R}}\left|
	\int_{\mathbb{R}}\frac{2yt}{(t^2+y^2)^2}\left[b(x-y)-b(x)\right]dy
	\right|^p dx
	\right).
\end{equation}
After an application of the H\"older's inequality to the internal integral we get:
\begin{multline}\label{lp430}
 \int_{\mathbb{R}} \frac{t^{1/q}}{(t^2+y^2)^{1/q}}\frac{2yt^{1/p}}{(y^2+t^2)^{2-1/q}}\left[b(x-y)-b(x)\right]dy \leq\\
  \leq \left(\int_\mathbb{R}\frac{t}{(t^2+y^2)}dy\right)^{1/q}
  \left(\int_\mathbb{R}\frac{2y^pt}{(y^2+t^2)^{1+p}}\left|b(x-y)-b(x)\right|^pdy\right)^{1/p}
\end{multline}
\begin{equation}\label{lp440}
	\leq \pi^{1/q}\left(\int_\mathbb{R}\frac{2y^pt}{(y^2+t^2)^{1+p}}\left|b(x-y)-b(x)\right|^pdy\right)^{1/p}
\end{equation}
and thus
\begin{equation}\label{lp450}
	\|J_1\|_{L^p(\mathbb{R}\times\mathbb{R}_+)}^p\leq \pi^{p/q}
	\int_{\mathbb{R}_+}dt\int\limits_{\mathbb{R}\times\mathbb{R}}
	\frac{2y^pt}{(y^2+t^2)^{1+p}}\left|b(x-y)-b(x)\right|^pdydx.
\end{equation}
Now since
\begin{equation}\label{lp460}
	\int_{\mathbb{R}_+}\frac{t}{(y^2+t^2)^{1+p}}dt = \frac{y^{-2p}}{2p},
\end{equation}
we get, using (\ref{not10}):
\begin{equation}\label{lp470}
	\|J_1\|_{L^p(\mathbb{R}\times\mathbb{R}_+)}^p \leq
	\frac{\pi^{p/q}}{p} \int\limits_{\mathbb{R}\times\mathbb{R}}
	\frac{\left|b(x-y)-b(x)\right|^p}{|y|^p}dydx
	\leq \frac{\pi^{p/q}}{p}\|b\|_{\dot W^{1-1/p}_p(\mathbb{R})}^p .
\end{equation}
Of course, since $b\in \dot W^{1-1/r}_r(\mathbb{R})$ for any $r\in (3/2, p]$ we actually have
\begin{equation}\label{lp485}
\|J_1\|_{L^r(\mathbb{R})} \leq \frac{\pi^{(r-1)/r}}{r^{1/r}}\|b\|_{\dot W^{1-1/r}_r(\mathbb{R})},
\end{equation}
which implies:
\begin{equation}
	\|\mathcal{F}^{-1}_k(I_1)\|_{L^r(\mathbb{R})} \leq 2^{1/2r}\frac{\pi^{(r-3/2)/r}}{r^{1/r}}\|b\|_{\dot W^{1-1/r}_r(\mathbb{R})},
\end{equation}
This type of calculations are known since the famous papers by Agmon, Douglis and Nirenberg (see \cite{ADN1}, \cite{ADN2}).

To finish our estimates for the gradient of the pressure we must deal with the term $I_2$ defined by (\ref{lp295}).
Estimates differ depending on the sign of $a_2$. We will be interested in a behaviour
of particular terms for $k\to 0$ and for $k\to\infty$, and we emphasize this by introducing a smooth cut-off function
$\zeta(k)$ such that $\zeta(k) = 1$ for $|k| \leq 1$ and $\zeta(k) = 0$ for $|k| > 1$. Then we split
integral $I_2$ as follows:
\begin{equation}\label{lp522_split}
	I_2(t, k) = \zeta(k) I_2 + (1-\zeta(k)) I_2 = I_{21}(t, k) + I_{22}(t, k),
\end{equation}
and estimate it separately.

First, let $a_2 > 0$. In this case we have 
\begin{equation}
\Re (a_2 + 2|k| + \sqrt{a_2^2+4(k^2 + a_1ik)}) > a_2 > 0	
\end{equation}
and
$\Re (\lambda_3(-f+\lambda_3)+k^2) \sim k^2$ for small $|k|$, thus we present $I_{21}$ as:
\begin{equation}\label{lp522_21}
	I_{21}(t, k) = e^{t\lambda_1}ik \varphi_{21}(k) (ik\hat{d}(k)),
\end{equation}
where
\begin{equation}
	\varphi_{21}(k) = \zeta(k)\frac{(a_2+2|k|+\sqrt{a_2^2+4(k^2+a_1ik)})\sigma(k)(\lambda_3(-f+\lambda_3)+k^2)}{-k^2(-f+\lambda_3+\lambda_1)}.
\end{equation}
It is then straightforward reasoning that a function $\varphi_{21}(k)$ is a proper multiplier in the sense of the Marcinkiewicz theorem.
Moreover, since $d\in \dot W^{2-1/p}_p(\mathbb{R})$ then $\mathcal{F}^{-1}_k(ik \hat d(k)) \in \dot W^{1-1/p}_p(\mathbb{R})$,
and we reuse techniques exploited earlier to estimate terms connected with $b(k)$, to obtain:
\begin{equation}
	\|\mathcal{F}^{-1}_{k}\left(I_{21}(t, k)\right)\|_{L^p(\mathbb{R}^2_+)} \leq C
		\|\mathcal{F}^{-1}_{k}\left(ik\hat{d}(k)\right)\|_{\dot W^{1-1/p}_p(\mathbb{R})} =
			\|d\|_{\dot W^{2-1/p}_p(\mathbb{R})}.
\end{equation}
Estimate of $I_{22}(t, k)$ similar to the previous one with one additional feature, that it does not depend on the sign of $a_2$.
We thus have:
\begin{equation}
	\|\mathcal{F}^{-1}_{k}\left(I_{22}(t, k)\right)\|_{L^p(\mathbb{R}^2_+)} \leq C
		\|\mathcal{F}^{-1}_{k}\left(ik\hat{d}(k)\right)\|_{\dot W^{1-1/p}_p(\mathbb{R})} =
			C\|d\|_{\dot W^{2-1/p}_p(\mathbb{R})}.	
\end{equation}

We would like to emphasize, that the estimate of $I_{22}$ does not depend on the sign of $a_2$, and hence
we can use it again in cases $a_2 = 0$ and $a_2 > 0$, provided $d\in \dot W^{2-1/p}_p(\mathbb{R})$.

Let us now consider the case $a_2 = 0$. We have 
\begin{equation}
2|k|+\sqrt{4(k^2+a_1ik)} \sim \sqrt{k},
\end{equation}
but also
\begin{equation}
(\lambda_3(-f+\lambda_3) + k^2) \sim \sqrt{k} \textrm{~for small~} |k|.	
\end{equation}
 This does not allow us to present
$I_{21}$ in the form (\ref{lp522_21}), but the following one:
\begin{equation}
	I_{21}(t, k) = e^{t\lambda_1} ik \varphi_{21}(k) \hat{d}(k),
\end{equation}
where $\varphi_{21}(k) = \zeta(k)\frac{(2|k|+\sqrt{4(k^2+a_1ik)})\sigma(k)(\lambda_3(-f+\lambda_3)+k^2)}{-ik(-f+\lambda_3+\lambda_1)}$
is a valid multiplier in the sense of the Marcinkiewicz theorem. Hence, a proper estimate is the following:
\begin{equation}
	\|\mathcal{F}^{-1}_{k}\left(I_{21}(t, k)\right)\|_{L^p(\mathbb{R}^2_+)} \leq C
			\|d\|_{\dot W^{1-1/p}_p(\mathbb{R})}.
\end{equation}
Thus, in the case $a_2 = 0$ integral $I_2$ can be estimated as follows:
\begin{equation}
	\|\mathcal{F}^{-1}_{k}\left(I_2\right)\|_{L^p(\mathbb{R}^2_+)} \leq
		C\|d\|_{\dot W^{2-1/p}_p(\mathbb{R})\cap \dot W^{1-1/p}_p(\mathbb{R})}.
\end{equation}

Let us now assume that $a_2 < 0$. In this case the term $(\lambda_3(-f+\lambda_3)+k^2) < a_2 < 0$, however
$a_2+2|k| + \sqrt{a_2^2+4(k^2+a_1ik)} \sim |k|$, and thus we can present term $I_{21}$ in the form
\begin{equation}
	I_{21}(t, k) = e^{t\lambda_1} ik \varphi_{21}(k) \hat{d}(k),
\end{equation}
where $\varphi_{21}(k) = \zeta(k)\frac{(2|k|+\sqrt{4(k^2+a_1ik)})\sigma(k)(\lambda_3(-f+\lambda_3)+k^2)}{-ik(-f+\lambda_3+\lambda_1)}$
is a proper multiplier in the sense of the Marcinkiewicz theorem, and we obtain:
\begin{equation}
	\|\mathcal{F}^{-1}_{k}\left(I_{21}(t, k)\right)\|_{L^p(\mathbb{R}^2_+)} \leq C
			\|d\|_{\dot W^{1-1/p}_p(\mathbb{R})},
\end{equation}
which, together with the standard estimate of $I_{22}$ gives us:
\begin{equation}
	\|\mathcal{F}^{-1}_{k}\left(I_2\right)\|_{L^p(\mathbb{R}^2_+)} \leq
		\|d\|_{\dot W^{2-1/p}_p(\mathbb{R})\cap \dot W^{1-1/p}_p(\mathbb{R})}.
\end{equation}

Gathering all above estimates we have proved the following inequalities:
\begin{itemize}
  	\item for $a_2 \leq 0$ and all $r\in (3/2, p]$: 
	  \begin{equation}
	  	\|\nabla p\|_{L^r(\mathbb{R}^2_+)} 
	  		\leq C(f, a_1, a_2)\left(\|b\|_{\dot W^{1-1/r}_r(\mathbb{R})} + 
	  					\|d\|_{\dot W^{2-1/r}_r(\mathbb{R})\cap \dot W^{1-1/r}_r(\mathbb{R})}\right).
	  \end{equation}
  	\item for $a_2 > 0$ and all $r\in (3/2, p]$:
	  \begin{equation}
	  	\|\nabla p\|_{L^r(\mathbb{R}^2_+)} 
	  		\leq C(f, a_1, a_2)\left(\|b\|_{\dot W^{1-1/r}_r(\mathbb{R})} + 
	  					\|d\|_{\dot W^{2-1/r}_r(\mathbb{R})}\right).
	  \end{equation}	
\end{itemize}
This completes the proof of Theorem \ref{R2pTheorem}.

\subsection{Second derivatives of the velocity -- reduction of the system.}\label{secondD}

In this section we introduce a homogeneous system for the velocity, from which 
it will be easier to obtain proper regularity of $\nabla^2 v$. Once we have
a simplified system we derive the solution and show estimates for it. 

First, let us recall that our solution to the system (\ref{lp0})-(\ref{lp30}) 
satisfies (independently of the sign of $a_2$) the following system:
\begin{equation}\label{lp550}
	\begin{array}{rcll}
	  	a_1 u_{,1} + a_2 u_{,2} - \Delta u & = & - \nabla p  \quad &\textrm{in~} \mathbb{R}^2_+,\\
	  	u & = & D & \textrm{on~}\partial\mathbb{R}^2_+,
 	\end{array}	
\end{equation}
where $D$ is in a proper class, which depends on the sign of $a_2$, namely:
\begin{itemize}
	\item for $a_2 < 0$: $D \in W^{2-1/r}_r(\mathbb{R})$,
	\item for $a_2 = 0$: $D \in \dot W^{2-1/r}_r(\mathbb{R})\cap \dot W^{1-1/r}_r(\mathbb{R})$,
	\item for $a_2 > 0$: $D \in \dot W^{2-1/r}_r(\mathbb{R})\cap \dot W^{2-2/r}_r(\mathbb{R})$,
\end{itemize}
for all $r\in (3, p]$, where we emphasize, that in case $a_2 < 0$ we need the full norm, not
only an homogeneous one.

We subtract inhomogeneity from the right hand side of (\ref{lp550}) without
changing the regularity of boundary condition $D$ in each of the cases of signum of $a_2$.
To obtain this we use Theorem \ref{OR2_80} and Theorem \ref{traceTh}. 

We present the solution $u$ as
\begin{equation}\label{lp580}
	u = v + w,
\end{equation}
where $w$ is a truncation to the halfspace $\mathbb{R}^2_+$ of the solution to the system in the whole $\mathbb{R}^2$ space:
\begin{equation}\label{lp590}
	a_1 w_{,1} + a_2 w_{,2} - \Delta w = -\widetilde{\nabla p},
\end{equation}
where $\widetilde{\nabla p}$ on the right hand side stands for its standard extension on the whole $\mathbb{R}^2$ with
a preservation of its  norm. Theorem \ref{OR2_80}
guarantees that the solution exists, thus $v$:
\begin{equation}\label{lp595}
\begin{array}{rcll}
  a_1 v_{,1} + a_2 v_{,2} - \Delta v & = & 0 & \textrm{in~} \mathbb{R}^2_+,\\
  v & = & D-w = \widetilde{D}\quad& \textrm{on~} \partial\mathbb{R}^2_+.
\end{array}	
\end{equation}

The question is, does $\widetilde{D}$ have the same regularity as $D$. 
Since for $r > 3$ we have $3r/(3+r) > 3/2$ and we have $\nabla p \in L^s(\mathbb{R}^2_+)$ for all $s \in (3/2, p]$
we are in position to use Theorem \ref{OR2_80} and Lemma \ref{OR2_139} to get:
\begin{equation}
	\|\nabla w\|_{W^1_r(\mathbb{R}^2)} \leq \|\nabla p\|_{L^r(\mathbb{R}^2)\cap L^{3r/(3+r)}(\mathbb{R}^2)},
\end{equation}
for all $r\in (3, p]$ and
\begin{equation}
	\|w_{|x_2 = 0}\|_{\dot W^{2-1/r}_r(\mathbb{R}) \cap \dot W^{2-2/r}_r(\mathbb{R})} \leq \|\nabla p\|_{L^r(\mathbb{R}^2)\cap L^{3r/(3+r)}(\mathbb{R}^2)}.
\end{equation}

This implies in particular, that for $a_2 \geq 0$ subtraction of $w$ does not change the regularity of
boundary conditions, hence $\widetilde{D}$ has the same regularity as $D$.

For $a_2 < 0$ we have different behaviour of eigenvalues and
we may use Theorem \ref{traceTh}. Since $\widetilde{\nabla p} \in L^r(\mathbb{R}^2)$ for all $r \in (3/2, p]$ we get
\begin{equation}
	\|w_{|x_2 = 0}\|_{W^{2-1/r}_r(\mathbb{R})} \leq \|\nabla p\|_{L^r(\mathbb{R}^2)},
\end{equation}
which again implies that $\widetilde{D}$ has the same regularity as $D$, namely: $\widetilde{D}\in W^{2-1/r}_r(\mathbb{R})$.

Above considerations justify the following set of inequalities:
\begin{itemize}
	\item for $a_2 < 0$:
		\begin{equation}\label{boundIneq1}
			\|\widetilde{D}\|_{W^{2-1/r}_r(\mathbb{R})} \leq C\left(\|D\|_{W^{2-1/r}_r(\mathbb{R})} + \|w_{|x_2 = 0}\|_{W^{2-1/r}_r(\mathbb{R})}\right),
		\end{equation}
	\item for $a_2 = 0$:
		\begin{multline}\label{boundIneq2}
			\|\widetilde{D}\|_{\dot W^{2-1/r}_r(\mathbb{R})\cap \dot W^{1-1/r}_r(\mathbb{R})} \\
				\leq C\left(\|D\|_{\dot W^{2-1/r}_r(\mathbb{R})\cap \dot W^{1-1/r}_r(\mathbb{R})} + \|w_{|x_2 = 0}\|_{\dot W^{2-1/r}_r(\mathbb{R})\cap \dot W^{1-1/r}_r(\mathbb{R})}\right),
		\end{multline}
	\item for $a_2 > 0$:
		\begin{multline}\label{boundIneq3}
			\|\widetilde{D}\|_{\dot W^{2-1/r}_r(\mathbb{R})\cap \dot W^{2-2/r}_r(\mathbb{R})} \\
				\leq C\left(\|D\|_{\dot W^{2-1/r}_r(\mathbb{R})\cap \dot W^{2-2/r}_r(\mathbb{R})} + \|w_{|x_2 = 0}\|_{\dot W^{2-1/r}_r(\mathbb{R})\cap \dot W^{2-2/r}_r(\mathbb{R})}\right).
		\end{multline}	
\end{itemize}

\textbf{Derivation of solution.}
For the readibility we again denote $\widetilde{D}$ as $D$, since in a view of (\ref{boundIneq1})-(\ref{boundIneq3}) this does not affect any estimates.
We solve the system (\ref{lp595}) in the same manner as the one considered earlier -- 
first we apply Fourier Transform and then solve ordinary
differential equations with initial data coming from the boundary constraints. 

Taking the Fourier transform of ($\ref{lp595}_1$) we get:
\begin{equation}\label{lp680}
	a_1 ik v_l + a_2\dot{v_l} + k^2 v_l - \ddot{v_l} = 0,\qquad	\textrm{where~}l = 1, 2,
\end{equation}
where again we denoted $\partial_{x_2}$ as $\dot{}$.
We easily compute a solution:
\begin{equation}\label{lp710}
	v_l(t, k) = e^{t\lambda_-(k)} V_{0l}(k),
\qquad \textrm{where~}
	V_{0i}(k) = \hat{D}_i(k).
\end{equation}
and
\begin{equation}
\lambda_-(k) = \frac{1}{2}(a_2 - \sqrt{a_2^2 + 4(k^2 + a_1 ik)}).
\end{equation}

\noindent \textbf{Remark: } The behaviour of $\lambda_-(k)$ at $|k| \to 0$ and $|k| \to \infty$ will be
crucial for our considerations. It is straightforward, that $\lambda_-(k) \sim |k|$ for
large $|k|$ independently of $a_1$ and $a_2$, however its behaviour at $0$ changes depending
on $a_1$ and $a_2$, namely, for small $k$:
\begin{itemize}
	\item for $a_2 < 0$: $\Re \lambda_-(k) < a_2 < 0$,
	\item for $a_2 = 0$: $\Re \lambda_-(k) \sim -\sqrt{|k|}$,
	\item for $a_2 > 0$: $\Re \lambda_-(k) \sim -|k|^2$.
\end{itemize}
We emphasize that if $a_2 = 0$, then $a_1^2 = a_1^2 + a_2^2 = v_\infty^2 \neq 0$.

\subsection{Estimate of $u_{,11}$}\label{estU11}
We start with the estimate of $u_{i, 11}$, which brings down
to an estimate of
\begin{equation}\label{lp750}
	e^{t\lambda_-(k)} (ik)^2 \hat{D}_i(k).
\end{equation}
Since we consider multiple cases it is thus reasonable to present them in separate lemmas.

\begin{Lemma}
	Given $u_{i, 11}$ in the form (\ref{lp750}). Assuming $a_2 \leq 0$ and $\dot D\in W^{2-1/p}_p(\mathbb{R})$
	one has:
	\begin{equation}
		\|u_{i, 11}\|_{L^p(\mathbb{R}^2_+)} \leq C \|D\|_{\dot W^{2-1/p}_p(\mathbb{R})}.
	\end{equation}
\end{Lemma}

\begin{Proof}{}
We see that $ik\hat{D}_i(k)$ has a good regularity (namely \newline \mbox{$\mathcal{F}^{-1}_k(ik\hat{D}_i(k)) \in \dot W^{1-1/p}_p$}) to get $L_p$-estimates for this term
(repeating the procedure for the gradient $\nabla p$), however
we need to show that one can change $e^{-t\lambda_-(k)}$ into $e^{-t|k|}$. 
Since $a_2 \leq 0$ there exists a constant $c(a_2)$ such that 
$\lambda_-(k) + c(a_2)|k| \leq 0$ for all $k\in \mathbb{R}$ and 
thus the following multiplier is valid in the sense of the Marcinkiewicz theorem:
\begin{equation}\label{lp760}
	\varphi(k) = e^{t(\lambda_-(k)+c(a_2)|k|)},
\end{equation}
and can be estimated independently of $t$.
Using it we are able to bring down estimate of $\|u_{i, 11}\|_{L^p(\mathbb{R}^2_+)}$ to estimate of a term
$\mathcal{F}^{-1}_{k}\left(-e^{-tc(a_2)|k|}ik (ik\hat{D}_i(k))\right)$, since
\begin{equation}
	\mathcal{F}_x (u_{i,11}) = e^{t(\lambda_-(k)+c(a_2)|k|)}\left(-e^{-tc(a_2)|k|}ik (ik\hat{D}_i(k))\right).
\end{equation}
Thus, in  case $a_2 \leq 0$,
\begin{multline}\label{lp780}
	\|u_{i, 11}\|_{L^p(\mathbb{R}^2_+)}
	 \leq\\
	 \leq C\|\mathcal{F}^{-1}_{k}\left(e^{-tc(a_2)|k|}ik (ik\hat{D}_i(k))\right)\|_{L^p(\mathbb{R}^2_+)}
	 \leq C(a_2)\|{D}_i(k)\|_{\dot W^{2-1/p}_p(\mathbb{R})}.
\end{multline}
\end{Proof}

We estimate term $u_{i, 11}$ for the case $a_2 > 0$ using the following Lemma:

\begin{Lemma}
	Given $u_{i, 11}$ in the form (\ref{lp750}). Assuming $a_2 > 0$ and $D\in \dot W^{2-1/p}_p(\mathbb{R})\cap \dot W^{2-2/p}_p(\mathbb{R})$
	one has:
	\begin{equation}
		\|u_{i, 11}\|_{L^p(\mathbb{R}^2_+)} \leq C \|D\|_{\dot W^{2-1/p}_p(\mathbb{R})\cap \dot W^{2-2/p}_p(\mathbb{R})}
	\end{equation}
\end{Lemma}

\begin{Proof}{}
Since the behaviour of $\lambda_-(k)$ is different
in a neighbourhood of $0$ and in a neighbourhood of $\infty$ we cannot use the technique use in the previous proof,
because there exists no constant $c(a_2)$ such that $\varphi(k)$ from (\ref{lp760}) is a valid multiplier.
We thus introduce a cut-off function $\pi(k)$ as follows:
\begin{equation}\label{lp790}
	\pi(k) \equiv 1 \quad \textrm{for~}|k|\leq L,\qquad \pi(k)\equiv 0\quad \textrm{for~}|k|\geq L+1,
\end{equation}
for some positive constant $L$, which we describe later. 

\noindent We split our term $e^{t\lambda_-(k)}ik (ik\hat{D}_i(k))$ as
\begin{equation}\label{lp800}
	e^{t\lambda_-(k)}ik \pi(k)(ik\hat{D}_i(k)) + e^{t\lambda_-(k)}ik (1-\pi(k))(ik\hat{D}_i(k)) =: I_1 + I_2.
\end{equation}
Let us first estimate $I_1$. We consider here the worst case, when $a_1 = 0$. All futher
estimates can be repeated for $a_1 \neq 0$. From the basic properties of $\lambda_-(k)$ we
see that for a proper constant $\tilde L$ a multiplier $\varphi_1$:
\begin{equation}\label{lp810}
	\varphi_1(k) = e^{t(\lambda_-(k)+\tilde L k^2)}\pi(k)
\end{equation}
is a good multiplier for all $t\geq 0$ in the sense of the Marcinkiewicz Theorem, with a proper estimate
not dependent of $t$ (the case when $\lambda_-(k) + \tilde L k^2 \leq 0$ for all $k$).  Thus we may write
\begin{equation}\label{lp820}
	\|\mathcal{F}^{-1}_{k}\left(I_1(t, \cdot)\right)\|_{L^p(\mathbb{R})} \leq
	\left\|\mathcal{F}^{-1}_{k}\left(e^{-\tilde L k^2 t}ik(ik \pi(k)\hat{D}_i(k))\right)\right\|_{L^p(\mathbb{R})}.
\end{equation}
The constant $\tilde L$ does not affect any estimates, so for the readibility of the paper
we assume that $\tilde L = 1$. We also denote $ik\pi(k) \hat{D}_i(k)$ as $\hat{B}_i(k)$ and $J_1(t, x)$ as
\begin{equation}\label{lp825}
	J_1(t, x) = \mathcal{F}^{-1}_{k}\left(e^{-tk^2}ik\hat{B}_i(k)\right)
\end{equation}
Since
\begin{equation}\label{lp830}
	\mathcal{F}^{-1}_{k}\left(e^{-tk^2}\right) = \frac{e^{-x^2/4t}}{\sqrt{2}\sqrt{t}}
\end{equation}
we have:
\begin{equation}\label{lp840}
	\mathcal{F}^{-1}_{k}\left(e^{-tk^2}ik\right) = \frac{e^{-x^2/4t}x}{2\sqrt{2}t^{3/2}}.
\end{equation}
Above term is integrable and odd with respect to $x$ so we may write:
\begin{equation}
	J_1(t, x) = \int_{\mathbb{R}} e^{-y^2/4t}\frac{y}{2\sqrt{2}t^{3/2}}[{B}_i(x-y) - {B}_i(x)] dy.
\end{equation}
Using H\"older inequality we get:
\begin{multline}
 \|J_1(t, \cdot)\|_{L^p(\mathbb{R})}^p  = \int_\mathbb{R} \left|\int_\mathbb{R} e^{-y^2/4t}\frac{y}{t^{3/2}}[{B}_i(x-y)-{B}_i(x)] dy\right|^p dx \\
 \leq \frac{1}{2\sqrt{2}}\left(\int_\mathbb{R} e^{-qy^2/8t}\frac{1}{t^{1/2}}\right)^{1/q}
 \left(\int_\mathbb{R} e^{-py^2/8t}\frac{y^p}{t^{p+1/2}} |{B}_i(x-y)-{B}_i(x)|^p dy\right).
\end{multline}
Now since $\int_\mathbb{R} e^{-qy^2/8t}\frac{1}{t^{1/2}} dy = \frac{2\sqrt{2\pi}}{\sqrt{q}}$ and
\begin{equation}
	\int_0^\infty e^{-py^2/8t}\frac{y^p}{t^{p+1/2}} dt = 2^{3(p-1/2)}p^{p-1/2}\Gamma(p-1/2)|y|^{1-p} = c_1(p) |y|^{1-p}
\end{equation}
we can write:
\begin{equation}
	\|J_1(\cdot, \cdot)\|_{L^p(\mathbb{R}^2_+)}^p \leq 2^{3(p-1/2)}p^{p-1/2}\Gamma(p-1/2)\frac{\sqrt{\pi}}{\sqrt{q}}
	\int_{\mathbb{R}\times\mathbb{R}}\frac{|{B}_i(x-y)-{B}_i(x)|^p}{|y|^{p-1}},
\end{equation}
where the right hand side can be estimated (right from the definition (\ref{not10})) by
\begin{equation}\label{lp850}
	2^{3(p-1/2)}p^{p-1/2}\Gamma(p-1/2)\frac{\sqrt{\pi}}{\sqrt{q}}\|{B}_i\|_{\dot W^{1-2/p}(\mathbb{R})}^p.
\end{equation}
This term, however, can be estimated by $C(q)\|{D}_i\|_{\dot W^{2-2/p}_p(\mathbb{R})}^p$,
since multiplication by a smooth function $\pi(k)$ does not change the class of the function $ik\hat{D}_i$ and
$\|\mathcal{F}^{-1}_k(ik\hat{D}_i)\|_{\dot W^{1-2/p}_p(\mathbb{R})} = \|\mathcal{F}^{-1}_k(\hat{D})_i\|_{\dot W^{2-2/p}_p(\mathbb{R})}$.

Before we make futher estimates we would like to emphasize, that this type of estimates and 
apperance of terms like (\ref{lp830}) are characteristic to a parabolic problem. 
We see, that a change of a sign of the coefficient $a_2$ results in the different behaviour of the
eigenvalue, which brings in this parabolic disturbance to our estimates and might be 
the cause of the presence of the wake region behind the obstacle.

Let us now return to the second term from (\ref{lp800}), i.e. $I_2$:
\begin{equation}\label{lp870}
	I_2 = e^{t\lambda_-(k)} ik (1-\pi(k)) (ik \hat{D}_i(k)).
\end{equation}
In this case we introduce a multiplier $\varphi(k) = e^{t(\lambda_-(k)+|k|/2)}$.
Since $\lambda_-(k)\sim -|k|$ for large $|k|$ we see, that
\begin{equation}\label{lp890}
	\lambda_-(k)+|k|/2 \sim -|k|/2 < 0 \qquad \textrm{for~}|k|\textrm{~large enough}. 
\end{equation}
Now we may go back to the definition of function $\pi$, i.e. (\ref{lp790}), and
set $L$ large enough (and in fact also $\widetilde{L}$ small enough) to ensure, that for $|k|>L+1$ inequality (\ref{lp890}) holds. Then
our multiplier can be estimated independently of $t$. Summing up:
\begin{equation}
	\|\mathcal{F}^{-1}_{k}\left(I_2(t, \cdot)\right)\|_{L^p(\mathbb{R})}\leq 
		\left\|\mathcal{F}^{-1}_{k}\left(e^{-t|k|/2}ik(1-\pi(k))(ik\hat{D}_i(k))\right)\right\|_{L^p(\mathbb{R})}
\end{equation}
and all estimates for the gradient of $p$ can be applied directly for this term, since
\begin{equation}\label{lp900}
	\mathcal{F}^{-1}_{k}\left(ik(1-\pi(k))\hat{D}_i(k)\right)\in \dot W^{1-1/p}_p(\mathbb{R}).
\end{equation}
This estimate completes the case of $u_{,11}$.
	
\end{Proof}

\textbf{Remark:} In above lemmas we used an assumption that ${D}_i\in \dot W^{2-1/p}_p(\mathbb{R})\cap\dot W^{2-2/p}_p(\mathbb{R})$,
but  in fact ${D}_i\in \dot W^{2-1/r}_r(\mathbb{R})\cap\dot W^{2-2/r}_r(\mathbb{R})$ for all $r\in (3, p]$ and thus our estimate
also holds for $\|\nabla^2 u\|_{L^r(\mathbb{R}^2_+)}$.

\subsection{Estimate of $u_{,22}$.}\label{estU22}
To complete the proof of Theorem \ref{R2pTheoremV} we now estimate $u_{,22}$, which corresponds to estimate of
\begin{equation}\label{lp905}
u_{,22} = \mathcal{F}^{-1}_{k}(v_{,22}(k, t)) = \mathcal{F}^{-1}_{k}\left(\lambda_-^2(k) e^{t\lambda_-(k)}\hat{D}_i(k)\right).
\end{equation}
Again, we treat all cases of signum of $a_2$ in a separate Lemma. The first one will be for the case $a_2 < 0$:

\begin{Lemma}\label{lp907}
	Given $u_{i, 22}$ in the form (\ref{lp905}). Assuming $a_2 < 0$ and $D\in W^{2-1/p}_p(\mathbb{R})$ one has the following
	inequality:
	\begin{equation}
		\|u_{i, 22}\|_{L^p(\mathbb{R}^2_+)} \leq C\|D\|_{W^{2-1/p}_p(\mathbb{R})}.
	\end{equation}
\end{Lemma}

\begin{Proof}{}
The problem one encounters is that for $a_2 < 0$ we have $\lambda_-^2(k) \geq a_2^2$, which obviously 
does not behave like $|k|^2$ for small $k$ and hence we cannot
write this term in a form like in (\ref{lp750}), that is why a different
approach is needed and we will investigate the case $a_2 < 0$ more thoroughly.

As usual we introduce a smooth cut off function $\pi(k)$ such that $\pi(k) \equiv 1$ for $|k| \leq L$, for some constant
$L > 0$, which will described later, and $\pi(k) \equiv 0$ for $|k| \geq L+1$. As we have seen many times,
multiplication by a smooth bounded function of compact support does not influence essential estimates. Keeping this in mind
we may write:
\begin{multline}\label{lp910}
	v_{i,11}(k, t) = c(a_2)e^{-t c(a_2)}\pi(k) \hat{D}_i(k) + |k|^2 e^{-t|k| c(a_2)}(1-\pi(k)) \hat{D}_i(k) \\=: I_1(t, k) + I_2(t, k)
\end{multline}
where a constant $c(a_2)$ may differ from one occurence to another.

Integral $I_1$ is easy to estimate, since ${D} \in W^{2-1/p}_p(\mathbb{R})$ for $a_2 < 0$
(and of course $\mathcal{F}^{-1}_k(\pi(k)\hat{D}) \in W^{2-1/p}_p(\mathbb{R})$), 
and in particular ${D}\in L^p(\mathbb{R})$, which gives us:
\begin{eqnarray}
	\|\mathcal{F}^{-1}_k(I_1(k, t))\|_{L^p(\mathbb{R}^2_+)}^p & \leq & \int_0^\infty c(a_2) e^{-t c(a_2)} \|{D}\|_{L^p(\mathbb{R})}^p dt \\
		& \leq & c(a_2) \|{D}\|_{L^p(\mathbb{R})}^p \leq c(a_2)\|{D}\|_{W^{2-1/p}_p(\mathbb{R})}^p.
\end{eqnarray}
Integral $I_2(t, k)$ can be estimated in the same way as it was made in case of $u_{,11}$, i.e. one presents $I_2(t, k)$ as
\begin{equation}\label{lp930}
	I_2(t, k) = -e^{-t|k|c(a_2)} ik \left(ik (1-\pi(k))\hat{D}_i(k)\right),
\end{equation}
and estimates as follows:
\begin{equation}\label{lp940}
	\|\mathcal{F}^{-1}_k(I_2(t, k))\|_{L^p(\mathbb{R}^2_+)} \leq C\|{D}_i\|_{\dot W^{2-1/p}_p(\mathbb{R})},
\end{equation}
thus
\begin{equation}
	\|v_{i, 11}\|_{L^p(\mathbb{R}^2_+)} \leq C\|D_i\|_{W^{2-1/p}_p(\mathbb{R})},
\end{equation}
and the proof of Lemma \ref{lp907} is complete.
\end{Proof}

For the case of $a_2 = 0$ we have the following lemma:
\begin{Lemma}
	Given $u_{i, 22}$ in the form (\ref{lp905}). Assuming $a_2 = 0$ and $D\in \dot W^{2-1/p}_p(\mathbb{R})\cap \dot W^{1-1/p}_p(\mathbb{R})$ one has the following
	inequality:
	\begin{equation}
		\|u_{i, 22}\|_{L^p(\mathbb{R}^2_+)} \leq C\|D\|_{\dot W^{2-1/p}_p(\mathbb{R})\cap \dot W^{1-1/p}_p(\mathbb{R})}.
	\end{equation}
\end{Lemma}

\begin{Proof}{}
In case $a_2 = 0$ one has $\lambda_-^2 \sim k^2 + aik = ik(a-ik)$ for small $k$ (we will treat this term as a part of derivative, i.e. $ik$,
and part of a multiplicator, i.e. $a-ik$) and thus, proceeding as earlier (introducing a cut-off function $\pi(k)$):
\begin{multline}\label{lp950}
	v_{i, 11}(k, t) = e^{-t\sqrt{|k|}}ik(-ik + a)\pi(k)\hat{D}_i(k) - e^{-t|k|c(a_2)}ik(1-\pi(k))ik\hat{D}_i(k) =\\
	=: I_1(t, k) + I_2(t, k).
\end{multline}
Integrals like $I_2(t, k)$ we have already seen how to estimate -- since $\mathcal{F}^{-1}_k(ik(1-\pi(k))\hat{D}_i(k)) \in \dot W^{1-1/p}_p(\mathbb{R})$
we get:
\begin{equation}\label{lp960}
	\|\mathcal{F}^{-1}_k(I_2(t, k))\|_{L^p(\mathbb{R}^2_+)} \leq C \|{D}_i\|_{\dot W^{2-1/p}_p(\mathbb{R})}.
\end{equation}
To estimate $I_1(t, k)$ we notice, that since there exists a constant $c_{a_1}$ such that
$-\sqrt{|k|} + c_{a_1}|k| \leq 0$ for small $k$, we may use Marcinkiewicz theorem for a multiplier
$\varphi(k) = \pi(k)e^{t(|k|-\sqrt{|k|})}$ to get that
\begin{multline}\label{lp980}
	\|\mathcal{F}^{-1}_k(I_1(t, k))\|_{L^p(\mathbb{R}^2_+)} \leq \|\mathcal{F}^{-1}_k(e^{-t|k|c(a_2)})ik (a-ik)\pi(k)\hat{D}_i(k)\|_{L^p(\mathbb{R}^2_+)} \\
		\leq C\|D\|_{\dot W^{1-1/p}_p(\mathbb{R})}.
\end{multline}
\end{Proof}

For the case of $a_2 > 0$ we have the following lemma:
\begin{Lemma}
	Given $u_{i, 22}$ in the form (\ref{lp905}). Assuming $a_2 > 0$ and $D\in \dot W^{2-1/p}_p(\mathbb{R})\cap \dot W^{2-2/p}_p(\mathbb{R})$ one has the following
	inequality:
	\begin{equation}
		\|u_{i, 22}\|_{L^p(\mathbb{R}^2_+)} \leq C\|D\|_{\dot W^{2-1/p}_p(\mathbb{R})\cap \dot W^{2-2/p}_p(\mathbb{R})}
	\end{equation}
\end{Lemma}

\begin{Proof}{}	
To estimate $u_{,22}$ for $a_2 > 0$ we proceed as earlier (introducing a cut-off function $\pi(k)$): 
since $\Re \lambda_-(k) \sim -|k|^2$ for small $|k|$ we may write $v_{,22}$ as follows:
\begin{multline}\label{lp1000}
	v_{i,22}(t, k) = -ik e^{-t|k|^2} \pi(k)|k|^2 ik\hat{\widetilde{D}}_i(k) - ik e^{-t|k|c(a_2)} (1-\pi(k))ik\hat{\widetilde{D}}_i(k) \\
	=: I_1(t, k) + I_2(t, k).
\end{multline}
Integral $I_2(t, k)$ can be estimated as follows:
\begin{equation}\label{lp1010}
	\|\mathcal{F}^{-1}_k (I_2(t, k))\|_{L^p(\mathbb{R}^2_+)} \leq C\|D\|_{\dot W^{2-1/p}_p(\mathbb{R})},
\end{equation}
while for $I_1(t, k)$ one has:
\begin{equation}\label{lp1020}
	\|\mathcal{F}^{-1}_k (I_1(t, k))\|_{L^p(\mathbb{R}^2_+)} \leq C\|D\|_{\dot W^{2-2/p}_p(\mathbb{R})},
\end{equation}
repeating estimates for $u_{,11}$ and keeping in mind, that $\pi(k)|k|^2$ is a proper multiplier in the sense of the Marcinkiewicz theorem,
since $\pi(k)$ has bounded support.

These estimates prove the following inequality:
\begin{equation}
	\|v_{i, 22}\|_{L^p(\mathbb{R}^2_+)} \leq C\|D\|_{\dot W^{2-1/p}_p(\mathbb{R})\cap \dot W^{2-2/p}_p(\mathbb{R})},
\end{equation}
which completes the proof of this lemma.
\end{Proof}

\textbf{Remark: } As was the case for $u_{i, 11}$ -- since $D(x)$ is in a family of spaces, i.e. not only for $p$ but also for all $r\in (3, p]$,
all above estimates are valid also for $\|u_{i, 22}\|_{L^r(\mathbb{R}^2_+)}$. This completes the proof of Theorem \ref{R2pTheoremV}.

\subsection{Summary.}

As we have seen in the proof of Theorem \ref{R2pTheoremV} different regularity of boundary condition
is needed in case of $u_{,11}$ and $u_{,22}$, however for the readability of the paper we did not differentiated it in
the statement of the theorem, however now we can set together all these requirements.
The following array shows, what regularity on ${D}$ is required in particular cases:
$$
\begin{array}{||c|c|c|c||}
	\hline\hline
	& a_2 < 0 & a_2 = 0 & a_2 > 0 \\
	\hline
	~v_{,11}~ & \quad{D}\in \dot W^{2-1/p}_p\quad& {D}\in \dot W^{2-1/p}_p & {D}\in \dot W^{2-1/p}_p\cap \dot W^{2-2/p}_p \\
	\hline
	v_{,22} & {D}\in W^{2-1/p}_p & \quad{D}\in \dot W^{2-1/p}_p \cap\dot W^{1-1/p}_p\quad& \quad{D}\in \dot W^{2-1/p}_p\cap \dot W^{2-2/p}_p\quad\\
	\hline\hline
\end{array}
$$

In this table the Reader can see, what is the connection between the class of regularity
for the boundary conditions and the sign of $a_2$, which corresponds to the type of
points on the boundary (i.e. type E (elliptic) for $a_2 < 0$, type $P$ (parabolic) for $a_2 > 0$
 and type $S$ for $a_2 = 0$). In front of the obstacle it is required that
 the boundary conditions are in the inhomogeneous class $W^{2-1/p}_p(\mathbb{R})$,
 a typical for a strongly elliptic problems. The situation behind the obstacle $a_2 > 0$
 appears to have also a parabolic disturbance, which can be seen by the need
 of the space $\dot W^{2-2/p}_p(\mathbb{R})$. Such class of regularity corresponds
 to the trace space for the standard heat equation.

\section{The system in the whole space $\mathbb{R}^2$}\label{sect:OR2}

In this part we would like to present results, which were used in the previous section.

The standard approach to whole space linear problems is the technique of the Fourier transform 
together with a multiplier theorem, for example Lizorkin Theorem (see Theorem \ref{Lizorkin}).
Using it are able to show the following theorems. We would like to mention
that for our purposes not all estimates in this theorem are needed. Some of them
are however necessary to show existence of solutions to the Navier-Stokes system 
(\ref{intr_NS_0})-(\ref{intr_NS_30}) in an exterior domain that is why we state
them and give a proof of some of them. If the Reader is interested in this problem
we refer him to \cite{Galdi} and \cite{PokornyPHD}.

\begin{Theorem}\label{OR2_80}
  	Let $F \in L^q(\mathbb{R}^2)$ and $1 < q < \infty$. Then there exists
	a solution $u = (u_1, u_2)$ to the system:
	\begin{eqnarray}\label{OR2_90}
	 a_1 u_{,1} + a_2 u_{,2} - \Delta u & = & F \qquad \textrm{in~} \mathbb{R}^2,
	\end{eqnarray}
	for which the following inequality holds:
	\begin{equation}\label{OR2_100}
		\|\nabla^2 u\|_{L^q(\mathbb{R}^2)} \leq C\|F\|_{L^q(\mathbb{R}^2)}.
	\end{equation}
	If $q < 3$ then also the following inequality holds:
	\begin{equation}\label{OR2_103}
		\|\nabla u\|_{L^{3q/(3-q)}(\mathbb{R}^2)} \leq C\|F\|_{L^q(\mathbb{R}^2)}.
	\end{equation}
	Moreover, as a direct result of  previous statements,
	if $q > 3$ and $F\in L^s(\mathbb{R}^2)$ for all $s\in (3/2, q]$, then
	for all $r\in (3, q]$
	\begin{equation}\label{OR2_105}
		\|\nabla u\|_{W^{1}_r(\mathbb{R}^2)} \leq C(r) \|F\|_{L^r(\mathbb{R}^2)\cap L^{3r/(3+r)}(\mathbb{R}^2)}.
	\end{equation}
\end{Theorem}

\begin{Proof}{}
	After rotating the coordinate system this problem corresponds to the problem
	\begin{equation}\label{OR2_130}
		\lambda u_1 - \Delta u = F \qquad \textrm{in~} \mathbb{R}^2.
	\end{equation}
	After applying the Fourier transform to the above equation and gets:
	\begin{equation}
		\hat{u}(\xi) = \frac{\hat{F}(\xi)}{i\xi_1 + |\xi|^2}.
	\end{equation}
	Using Theorem \ref{Lizorkin} one immediately gets (\ref{OR2_100}), since
	a multiplier $\frac{-\xi_i \xi_j}{i\xi_1 + |\xi|^2}$, which stands
	for a derivative $u_{,ij}$, is a proper bounded multiplier.
	
	To show (\ref{OR2_103}) we again use Theorem \ref{Lizorkin} with $\beta = 1/3$. We must show that the multiplier
	\begin{equation}
		\frac{|\xi_1|^{4/3}|\xi_2|^{1/3} + |\xi_2|^{4/3}|\xi_1|^{1/3}}{|\xi|^2+\lambda|\xi_1|}
	\end{equation}
	is bounded for all $\xi\in \mathbb{R}^2$. Since
	\begin{equation}
		|\xi_1|^{4/3}|\xi_2|^{1/3} + |\xi_2|^{4/3}|\xi_1|^{1/3} \leq |\xi_1|^{1/3}|\xi|^{4/3}
	\end{equation}
	and
	\begin{equation}
		(|\xi|^{2/3}+\lambda^{1/3}|\xi_1|^{1/3})^3 \leq C(|\xi|^2+\lambda|\xi_1|)
	\end{equation}
	we get
	\begin{equation}
		\lambda^{1/3}|\xi_1|^{1/3}|\xi|^{4/3} \leq |\xi|^2+\lambda|\xi_1|
	\end{equation}
	thus
	\begin{equation}
		\lambda^{1/3}\|\nabla u\|_{L^{3p/(3-p)}(\mathbb{R}^2)} \leq C\left(\|F\|_{L^p(\mathbb{R}^2)} + \|G\|_{W^{1, p}(\mathbb{R}^2)}\right).
	\end{equation}	
\end{Proof}

As a direct result of this theorem we have the following:
\begin{Lemma}\label{OR2_139}
	Given $3 < q < \infty$. If $F \in L^q(\mathbb{R}^2)\cap L^1(\mathbb{R}^2)$ then
	the solution $u = (u_1, u_2)$ to the system from Theorem \ref{OR2_80} satisfies the following estimates:
	\begin{equation}\label{OR2_140}
		\|u\|_{W^2_r(\mathbb{R}^2)} \leq C(\lambda, r) \|F\|_{L^r_{A_r}}
		\quad \textrm{for all~}r \in (3, q],
	\end{equation}
	where $A_r = \{3r/(3+2r), 3r/(3+r), r\}$,
	\begin{equation}\label{OR2_150}
		\|\nabla u\|_{W^1_r(\mathbb{R}^2)} \leq C(\lambda, r) \|F\|_{L^q_{B_r}}
		\quad \textrm{for all~}r \in (3/2, q],
	\end{equation}
	where $B_r = \{3r /(3+r), r\}$.
	
	\noindent Moreover for all $r\in (3/2, q]$ one has:
	\begin{equation}\label{OR2_160}
		u_{|x_2 = 0}\in \dot W^{2-1/r}_r(\mathbb{R}) \cap \dot W^{1-1/r}_r(\mathbb{R}).
	\end{equation}
\end{Lemma}

\begin{Proof}{}
	Let $3/2 < r \leq q$. We take $r_1 = 3r/(3+r)$ and $r_2 = 3r/(3+2r)$. 
	Using previous theorem we immediately get $\|\nabla^2 u\|_{L^r} \leq \|F\|_{L^r}$, since
	$\|F\|_{L^r}\leq C(r)\|F\|_{L^q\cap L^1}$.
	Since $F\in L^q\cap L^1$ and we also have $F\in L^{r_1}$ and $F\in L^{r_2}$. Since $r_1 < 3$ this implies
	that $\|\nabla u\|_{L^r} =	 \|\nabla u\|_{L^{3r_1/(3-r_1)}} \leq \|F\|_{L^{r_1}}$, which is the desired
	estimate (\ref{OR2_150}). The same thing we can make with $r_2$ and $u$, since $r_2 < 3/2$ and
	thus $\|u\|_{L^r} = \|u\|_{L^{3r_2/(3-2r_2)}} \leq \|F\|_{L^{r_2}}$ and hence the proof of (\ref{OR2_140}) and
	(\ref{OR2_150}) is complete.
	
	To show (\ref{OR2_160}) one must notice, that since $\nabla u_{|x_2 = 0}\in W^{1-1/r}_r(\mathbb{R})$
	we get $u_{|x_2 = 0} \in \dot W^{2-1/r}_r(\mathbb{R})$ (straightforward from definition (\ref{not10})).
	The fact that $u_{|x_2 = 0} \in \dot W^{1-1/r}_r(\mathbb{R})$ can be shown using Lemma \ref{W1-1p}
	for $s = 3+\epsilon$ and $m = r$.
		
	\textbf{Remark:} bounds from (\ref{OR2_140}) and (\ref{OR2_150}) come from the inequalities
	$3s/(3-s) > 3/2$ and $3s/(3-2s) > 3$ for all $s > 1$.
\end{Proof}

Using different techniques than those presented in the proof of Theorem \ref{OR2_80} we are able to show the following
result:
\begin{Theorem}\label{traceTh}
	Let $f \in L^p(\mathbb{R}^2)$ such that $\textrm{supp~}f \subset \mathbb{R}^2_+$. Given a solution to the following system:
	\begin{equation}\label{traceTh10}
		a_1 u_{,1} + a_2 u_{,2} - \Delta u = f \qquad \textrm{in~}\mathbb{R}^2,
	\end{equation}
	with a condition at infinity $|u|\to 0\qquad \textrm{as~}|x|\to\infty$.
	Provided $a_2 < 0$, the following estimate is valid:
	\begin{equation}\label{traceTh20}
		\|u_{|x_2 = 0}\|_{W^{2-1/p}_p(\mathbb{R})} \leq C \|f\|_{L^p(\mathbb{R}^2)}.
	\end{equation}	
\end{Theorem}

\begin{Proof}{}
	From Lemma \ref{OR2_139} we have immediately 
	\begin{equation}
	\|\nabla^2 u\|_{L^p(\mathbb{R}^2)} \leq \|f\|_{L^p(\mathbb{R}^2)},
	\end{equation}
	so to prove Theorem \ref{traceTh} we need to show only $L_p$-- estimate for the function $u$, namely we prove the following inequality:
	\begin{equation}
		\|u_{|x_2 = 0}\|_{L^p(\mathbb{R})} \leq C \|f\|_{L^p(\mathbb{R}^2)}.
	\end{equation}
	We apply the Fourier transform in $x_1$ variable to (\ref{traceTh10}) to obtain the following differential equation:
	\begin{equation}\label{traceTh30}
		a_1 i\xi v + a_2 \dot v + \xi^2 v - \ddot v = \hat f(t, \xi),
	\end{equation}
	where $v(\xi, t) = \mathcal{F}_{x_1}(u)(\xi, t)$, and we denoted $x_2$ coordinate as $t$.
	
	With this system two eigenvalues are connected: stable $\lambda_- = (a_2 - \Delta)/2$ and unstable $\lambda_+ = (a_2 + \Delta)/2$,
	where $\Delta = \sqrt{a_2^2 + 4(\xi^2+a_1i\xi)}$.
	Let us observe, that $\Re \lambda_- < a_2$, which will be crucial for our considerations.
	The solution satisfies the following equation:
	\begin{equation}
		v(\xi, t) = \int_{-\infty}^t \frac{1}{\Delta} e^{\lambda_+(s-t)}\hat f(s, \xi) ds +
			\int_{t}^\infty \frac{1}{\Delta} e^{\lambda_-(s-t)}\hat f(s, \xi) ds.
	\end{equation}
	Since the support of $\hat f$ is a subset of $\mathbb{R}_2^+$ we have
	\begin{equation}
		v(\xi, 0) = \int_{0}^\infty \frac{1}{\Delta} e^{\lambda_-s}\hat f(s, \xi) ds.
	\end{equation}
	To estimate $\|u_{|x_2 = 0}\|_{L^p(\mathbb{R})}$ we use Marcinkiewicz theorem, i.e.
	\begin{eqnarray}
		\|u_{|x_2 = 0}\|_{L^p(\mathbb{R})} & = &\|\mathcal{F}^{-1}_\xi (v(\xi, 0)) \|_{L^p(\mathbb{R})} =
			\left\|\mathcal{F}^{-1}_\xi \left(\int_{0}^\infty \frac{1}{\Delta} e^{\lambda_-s}\hat f(\xi, s) ds\right)\right\|_{L^p(\mathbb{R})} \\
			& \leq & \int_{0}^\infty \|\mathcal{F}^{-1}_\xi \left(\frac{1}{\Delta} e^{\lambda_-s}\hat f(\xi, s) \right)\|_{L^p(\mathbb{R})} \\
			& \leq & \int_{0}^\infty C_M(s) \|f(\cdot, s)\|_{L^p(\mathbb{R})} ds,
	\end{eqnarray}
	where the term $C_M(s)$ comes from the term $\frac{1}{\Delta} e^{\lambda_-s}$, which, for convenience, we denote as $\Psi(\xi, s)$.
	An estimate of the constant $C_M(s)$, which comes from from the Marcinkiewicz theorem, is crucial for our estimate.
	Since we are in one dimension the constant $C(s)$ is estimated by the term, which is strongly convergent to $0$, since $a_2 < 0$:
	\begin{equation}
		C_M(s) \leq \sup_{\xi \in \mathbb{R}\setminus\{0\}} (|\Psi(\xi, s)| + |\xi\partial_{\xi}\Psi(\xi, s)|) \leq Ce^{a_2s/4}.
	\end{equation}
	This implies, that $\|u\|_{L^p(\mathbb{R})}$ can be estimated as follows:
	\begin{equation}
		\|u_{|x_2 = 0}\|_{L^p(\mathbb{R})} \leq C \|f\|_{L^p(\mathbb{R}^2_+)},
	\end{equation}
	which is the desired estimate.
\end{Proof}

The following Theorem is well known (see \cite{Galdi}):
\begin{Theorem}\label{OR2_40}\textbf{Oseen system in the full space $\mathbb{R}^2$.}
	Let $F \in L^q(\mathbb{R}^2)$, $G\in W^{1, q}(\mathbb{R}^2)$ and $1 < q < \infty$. Then there exists
	a solution $u = (u_1, u_2)$ and $p$ to the following 
	inhomogeneous Oseen system:
	\begin{eqnarray}
	 \lambda u_{,1} - \Delta u + \nabla p & = & F,\\
	 \nabla\cdot u & = & G,
	\end{eqnarray}
	which satisfies the following estimates:
	\begin{itemize}
		\item for all $1 < q < \infty$:
			\begin{equation}
				\lambda \|\nabla u_2\|_{L^q} + \lambda\left\|\frac{\partial u}{\partial x_1}\right\|_{L^q}
				+ \|\nabla^2 u\|_{L^q} + \|\nabla p\|_{L^q} \leq c(\|F\|_{L^q} + \|G\|_{W^{1,q}}),
			\end{equation}
		\item 	for all $1 < q < 3$: $\lambda^{1/3} \|\nabla u\|_{L^{3q/(3-q)}} \leq c(\|F\|_{L^q} + \|G\|_{W^{1,q}}),$
		\item for all $1 < q < 3/2$: $\lambda^{2/3}\|u\|_{L^{3q/(3-2q)}} \leq c(\|F\|_{L^q} + \|G\|_{W^{1,q}}).$			
	\end{itemize}
\end{Theorem}

As a direct application of the above Theorem we have the following Lemma:
\begin{Lemma}\label{OR2_75}
	Given $3 < q < \infty$. If $F \in L^q(\mathbb{R}^2)\cap L^1(\mathbb{R}^2)$ and 
	$G \in W^1_q(\mathbb{R}^2)\cap W^1_1(\mathbb{R}^2)$ then
	the solution $u = (u_1, u_2)$ and $p$ to the system from Theorem \ref{OR2_40} satisfies the following estimates:
	\begin{equation}\label{OR2_76}
		\|u\|_{W^2_r(\mathbb{R}^2)} \leq C(\lambda, r) \left(\|F\|_{L^q\cap L^1} + \|G\|_{W^1_q\cap W^1_1}\right)
		\quad \textrm{for all~}r \in (3, q],
	\end{equation}
	\begin{equation}\label{OR2_77}
		\|\nabla u\|_{W^1_r(\mathbb{R}^2)} \leq C(\lambda, r) \left(\|F\|_{L^q\cap L^1} + \|G\|_{W^1_q\cap W^1_1}\right)
		\quad \textrm{for all~}r \in (3/2, q].	
	\end{equation}
	Moreover for all $r\in (3/2, q]$ one has:
	\begin{equation}\label{OR2_78}
		u_{|x_2 = 0}\in \dot W^{2-1/r}_r(\mathbb{R}) \cap \dot W^{1-1/r}_r(\mathbb{R}),
	\end{equation}
	and for all $r\in (3, q]$ one has:
	\begin{equation}\label{OR2_79}
		u_{|x_2 = 0}\in \dot W^{1-1/r}_r(\mathbb{R}) \cap \dot W^{1-2/r}_r(\mathbb{R}).
	\end{equation}
\end{Lemma}

\begin{Proof}{}
	The proof of this lemma is analogous to the proof of Lemma \ref{OR2_139}.
	Property (\ref{OR2_79}) is a direct consequence of (\ref{OR2_76}).
\end{Proof}

\section{Proof of Theorem \ref{mainTh}}


In this section we give a proof of Theorem \ref{mainTh}. We extensively use results for the whole space $\mathbb{R}^2$
and for the halfspace $\mathbb{R}^2_+$.

To prove Theorem \ref{mainTh} we use a standard approach.
We consider two auxiliary problems: one in the whole space and the 
second one in a bounded domain (some neighbourhood of the boundary of the original domain).
With the former we deal with in Section \ref{sect:OR2}. To solve the latter one may use 
the standard technique of partition of unity, namely, spliting a neighbourhood of the boundary
into parts $U_i$ small enough to introduce a proper curvilinear system in each of them. In
this curvilinear coordinates the original problem transforms into a similar problem in a halfspace.
Moreover -- the support of a corresponding solution is contained in $U_i$. 

Existence of solutions is assured thanks to our assumptions ($F\in H^{-1}(\Omega)$, etc.),
since then one may use standard techniques for Hilbert spaces. We refer the Reader to \cite{2dextlin},
where a similar linear problem is considered and using these results we are able to show
existence also for the Oseen system in an elementary way.

Once we have a solution we may use mentioned technique of partition of unity and show additional regularity.

Results in the full space $\mathbb{R}^2$ apply directly, however in the case of the halfspace $\mathbb{R}^2_+$ it 
cannot be made without an effort, since assumptions in the halfspace require that $p > 3/2$ in case of the pressure $p$
and $p > 3$ in case of the velocity $v$, however for Theorem \ref{mainTh} to be applicable as a tool to prove Theorem \ref{NSTh} one has to have
this type of results for $p < 6/5$.

We assume only, that $p > 1$. Since in applications we are interested in $p < 6/5$, we will focus on the case
$p \in (1, 3/2)$. Before we continue we would like to mention two simple but important properties: if $p\in (1, 3/2)$ then $3p/(3-p) \in (3/2, 3)$
and $3p/(3-2p) \in (3, \infty)$.

We start with estimates on $\nabla p$. Recalling the Remark to Theorem \ref{R2pTheorem} we know, that estimates on $\nabla p$
are valid not only for $p > 3/2$, but for all $p > 1$ -- the constraint $p > 3/2$ came from the fact, that we wanted
to remove inhomogeneity from the right hand side while keeping proper estimates on boundary conditions. A similar
condition $p > 3$ was necessary in case of the velocity $v$.

In this section we will not only use stated theorems and lemmas but we will go into the details of their proofs.

As was mentioned before, after a localization procedure we end up with system (\ref{lpa0})-(\ref{lpa30}), where
$F \in L^r(\mathbb{R}^2_+)$, $G\in W^{1}_r(\mathbb{R}^2_+)$, $\underline{b} \in W^{1-1/r}_r(\mathbb{R})$ and
$\underline{d}\in W^{2-1/r}_r(\mathbb{R})$ for all $r\in (1, p]$. 
The next step us to solve in a similary way to Lemma \ref{extLemma} an auxiliary system in the full space
$\mathbb{R}^2$ obtaining the solution $(\widetilde{v}, \widetilde{q})$. Of course, since $p < 3/2$ we
are not able to obtain the same conditions on traces of $v$ and $\nabla v$. Using Theorem \ref{OR2_40} we
get:
\begin{eqnarray}
	\vec{n}\cdot\mathbb{T}(\widetilde{v}, \widetilde{q})_{|x_2 = 0}\cdot\vec{\tau} & \in & \dot W^{1-1/r}_r(\mathbb{R}), \label{mainTh10}\\
	f(v\cdot\vec{\tau})_{|x_2 = 0} & \in & \dot W^{2-1/r}_r(\mathbb{R}) \cap \dot W^{1-r_1}_{r_1}(\mathbb{R}),\label{mainTh20}\\
	\vec{n}\cdot v_{|x_2 = 0} & \in & \dot W^{2-1/r}_r(\mathbb{R}) \cap \dot W^{1-r_1}_{r_1}(\mathbb{R})\label{mainTh30},
\end{eqnarray}
where $r_1 = 3r/(3-r)$
where two last properties come from the fact, that $\nabla^2 v\in L^r(\mathbb{R}^2_+$ and $\nabla v\in L^{r_1}(\mathbb{R}^2_+)$.

In such a case, a subtraction $u = v-\widetilde{v}$ and $p = q-\widetilde{q}$ implies that we obtain the system (\ref{lp0})-(\ref{lp30})
for $u$, but $b$ and $d$ are of different regularity, namely:
\begin{eqnarray}
	b & = & \underline{b} - \widetilde{b} \in W^{1-1/r}_r(\mathbb{R}) + \dot W^{1-1/r}_r(\mathbb{R}) + \dot W^{1-1/r_1}_{r_1}(\mathbb{R})\cap \dot W^{2-1/r}_r(\mathbb{R})\label{mainTh40}\\
	d & = & \underline{d} - \widetilde{d} \in W^{2-1/r}_r(\mathbb{R}) + \dot W^{1-1/r_1}_{r_1}(\mathbb{R})\cap \dot W^{2-1/r}_r(\mathbb{R}),\label{mainTh50}
\end{eqnarray}
where $r_1 = 3r/(3-r)$ and $r\in (1, p]$.
In the proof of Theorem \ref{R2pTheorem} we used an assumption $b\in \dot W^{1-1/r}_r(\mathbb{R})$ and we see, that (\ref{mainTh40}) is strong enough
to obtain the following inequality:
\begin{equation}\label{mainTh60}
	\|b\|_{\dot W^{1-1/r}_r(\mathbb{R}) + \dot W^{1-1/r_1}_{r_1}(\mathbb{R})} 
		\leq \|\underline{b}-\widetilde{b}\|_{W^{1-1/r}_r(\mathbb{R}) + \dot W^{1-1/r}_r(\mathbb{R}) + \dot W^{1-1/r_1}_{r_1}(\mathbb{R})\cap \dot W^{2-1/r}_r(\mathbb{R})}.
\end{equation}
In the case of $d$ we are able to derive from (\ref{mainTh50}) the following inequality:
\begin{multline}\label{mainTh70}
	\|d\|_{\dot W^{1-1/r_1}_{r_1}(\mathbb{R})\cap \dot W^{2-1/r}_{r}(\mathbb{R}) + \dot W^{1-1/r}_r(\mathbb{R})\cap \dot W^{2-1/r}_r(\mathbb{R})} 
		\\ \leq \|\underline{d}-\widetilde{d}\|_{W^{2-1/r}_r(\mathbb{R}) + \dot W^{1-1/r_1}_{r_1}(\mathbb{R})\cap \dot W^{2-1/r}_r(\mathbb{R})}.
\end{multline}
These two inequalities imply that $\nabla p \in L^r(\mathbb{R}^2_+) + L^{r_1}(\mathbb{R}^2_+)$ and the following inequality is valid:
\begin{multline}\label{mainTh75}
	\|\nabla p\|_{L^r(\mathbb{R}^2_+) + L^{r_1}(\mathbb{R}^2_+)} \leq
		C(\|b\|_{\dot W^{1-1/r}_r(\mathbb{R}) + \dot W^{1-1/r_1}_{r_1}(\mathbb{R})} + \\
		\|d\|_{\dot W^{1-1/r_1}_{r_1}(\mathbb{R})\cap \dot W^{2-1/r}_{r}(\mathbb{R}) + \dot W^{1-1/r}_r(\mathbb{R})\cap \dot W^{2-1/r}_r(\mathbb{R})}).
\end{multline}
Indeed, to see this result we need to go into the details of the proof of Theorem \ref{R2pTheorem}. Since our problem
is linear we may treat influence of $b$ and $d$ separately, say $p = p_b + p_d$. As we have mentioned before, during
an estimate of $p$ we used a seminorm $\|b\|_{\dot W^{1-1/r}_r(\mathbb{R})}$, that is why we present
$b$ as $b = b_1 + b_2$, where $b_1 \in \dot W^{1-1/r}_r(\mathbb{R})$ and $b_2\in \dot W^{1-1/r_1}_{r_1}(\mathbb{R})$ to get:
\begin{multline}
	\|\nabla p_b\|_{L^r(\mathbb{R}^2_+)+L^{r_1}(\mathbb{R}^2_+)} \leq
		\|\nabla p_{b_1}\|_{L^r(\mathbb{R}^2_+)} + \|\nabla p_{b_2}\|_{L^{r_1}(\mathbb{R}^2_+)}\\
			\leq \|b\|_{\dot W^{1-1/r}_r(\mathbb{R}) + \dot W^{1-1/r_1}_{r_1}(\mathbb{R})}.
\end{multline}
The case with $d$ is a little bit different. In the proof of Theorem \ref{R2pTheorem}, during the estimate
of $\nabla p$ connected with a term $d$ we splitted its Fourier transform into 
$I_{21}(t, k) + I_{22}(t, k)$ (see \ref{lp522_split}). The part $\mathcal{F}^{-1}_{k}\left(I_{21}\right)$
was estimated by $\|d\|_{\dot W^{2-1/p}_p(\mathbb{R})}$, and the part $\mathcal{F}^{-1}_{k}\left(I_{22}\right)$
was estimated by $\|d\|_{\dot W^{1-1/p}_p(\mathbb{R})}$. Now since in our case we have
\begin{equation}
	d \in \dot W^{1-1/r_1}_{r_1}(\mathbb{R})\cap \dot W^{2-1/r}_{r}(\mathbb{R}) + \dot W^{1-1/r}_r(\mathbb{R})\cap \dot W^{2-1/r}_r(\mathbb{R})
\end{equation}
this implies the following inequality:
\begin{multline}
	\|\nabla p\|_{L^{r}(\mathbb{R}^2_+) + L^{r_1}(\mathbb{R}^2_+)} \leq
		\|\mathcal{F}^{-1}_{k}\left(I_{21})\right)\|_{L^r(\mathbb{R}^2_+)} +
		\|\mathcal{F}^{-1}_{k}\left(I_{22}\right)\|_{L^{r_1}(\mathbb{R}^2_+)}\\
		\leq \|d\|_{\dot W^{1-1/r_1}_{r_1}(\mathbb{R})\cap \dot W^{2-1/r}_{r} + \dot W^{1-1/r}_r(\mathbb{R})\cap \dot W^{2-1/r}_r(\mathbb{R})}.
\end{multline}
These considerations justify (\ref{mainTh75}).

In an estimate of the velocity $v$ we need not only homogeneous norm, but also $L^q$-norms on the boundary. That is why we must
check, to which spaces our boundary conditions $\widetilde{b}$, $\widetilde{d}$ belong to.

Lemma \ref{mainTh100} together with Theorem \ref{OR2_40} (see also previous estimates (\ref{mainTh10})-(\ref{mainTh30})) give us the following properties:
\begin{eqnarray}
	\vec{n}\cdot\mathbb{T}(\widetilde{v}, \widetilde{q})_{|x_2 = 0}\cdot\vec{\tau} & \in & \dot W^{1-1/r}_r(\mathbb{R}) \cap (L^r(\mathbb{R})+L^{r_1}(\mathbb{R})), \label{mainTh110}\\
	f(v\cdot\vec{\tau})_{|x_2 = 0} & \in & \dot W^{2-1/r}_r(\mathbb{R}) \cap \dot W^{1-r_1}_{r_1}(\mathbb{R})\cap (L^{r_1}(\mathbb{R})+L^{r_2}(\mathbb{R})),\label{mainTh120}\\
	\vec{n}\cdot v_{|x_2 = 0} & \in & \dot W^{2-1/r}_r(\mathbb{R}) \cap \dot W^{1-r_1}_{r_1}(\mathbb{R})\cap (L^{r_1}(\mathbb{R})+L^{r_2}(\mathbb{R}))\label{mainTh130},
\end{eqnarray}
where $r_1 = 3r/(3-r)$ and $r_2 = 3r/(3-2r)$. Using this properties we have:
\begin{multline}\label{mainTh140}
	b = \underline{b} - \widetilde{b} \in W^{1-1/r}_r(\mathbb{R}) + \dot W^{1-1/r}_r(\mathbb{R})\cap (L^r(\mathbb{R})+L^{r_1}(\mathbb{R})) \\
		+ \dot W^{1-1/r_1}_{r_1}(\mathbb{R})\cap \dot W^{2-1/r}_r(\mathbb{R})\cap (L^{r_1}(\mathbb{R})+L^{r_2}(\mathbb{R}))
\end{multline}
and
\begin{multline}\label{mainTh150}
	d = \underline{d} - \widetilde{d} \in W^{2-1/r}_r(\mathbb{R}) + \dot W^{1-1/r_1}_{r_1}(\mathbb{R})\cap \dot W^{2-1/r}_r(\mathbb{R})\cap (L^{r_1}(\mathbb{R})+L^{r_2}(\mathbb{R})).
\end{multline}
For our purposes we will need the following inequalities, which are a consequence of the above properties:
\begin{multline}\label{mainTh160}
	\|b\|_{\dot W^{1-1/r}_r(\mathbb{R})\cap (L^r(\mathbb{R})+L^{r_1}(\mathbb{R})) + \dot W^{1-r_1}_{r_1}(\mathbb{R}) \cap (L^{r_1}(\mathbb{R})+L^{r_2}(\mathbb{R}))} \leq \\
		\|\underline{b}-\widetilde{b}\|_{W^{1-1/r}_r(\mathbb{R}) + \dot W^{1-1/r}_r(\mathbb{R})\cap (L^r(\mathbb{R})+L^{r_1}(\mathbb{R})) 
		+ \dot W^{1-1/r_1}_{r_1}(\mathbb{R})\cap \dot W^{2-1/r}_r(\mathbb{R})\cap (L^{r_1}(\mathbb{R})+L^{r_2}(\mathbb{R}))}
\end{multline}
and
\begin{multline}\label{mainTh170}
	\|d\|_{W^{2-1/r}_r(\mathbb{R}) + \dot W^{2-1/r}_r(\mathbb{R}) \cap \dot W^{1-1/r_1}_{r_1}(\mathbb{R}) \cap (L^{r_1}(\mathbb{R})+L^{r_2}(\mathbb{R}))} \leq \\
		\|\underline{d}-\widetilde{d}\|_{W^{2-1/r}_r(\mathbb{R}) + \dot W^{1-1/r_1}_{r_1}(\mathbb{R})\cap \dot W^{2-1/r}_r(\mathbb{R})\cap (L^{r_1}(\mathbb{R})+L^{r_2}(\mathbb{R}))}.
\end{multline}

We are now in position to use Lemma \ref{boundLemma} to derive proper class for Dirichlet boundary conditions.
Using (\ref{mainTh160}) and (\ref{mainTh170}) it is not hard to see, that:
\begin{multline}\label{mainTh180}
	D(x_1) \in (L^{r_1}(\mathbb{R}) + L^{r_2}(\mathbb{R}))\cap \dot W^{1-1/r_1}_{r_1}(\mathbb{R}) \cap \dot W^{2-1/r}_r(\mathbb{R}) + W^{2-1/r}_r(\mathbb{R}) \\+
		(L^{r_1}(\mathbb{R})+L^{r_2}(\mathbb{R}))\cap \dot W^{1-1/r_1}_{r_1}(\mathbb{R})\cap \dot W^{2-1/r_1}_{r_1}(\mathbb{R})+\\
		(L^{r}(\mathbb{R})+L^{r_1}(\mathbb{R}))\cap \dot W^{1-1/r}_{r}(\mathbb{R})\cap \dot W^{2-1/r}_{r}(\mathbb{R}).
\end{multline}
The important thing in the space from (\ref{mainTh180}) is that it is a sum of spaces of a particular form:
\begin{equation}\label{mainTh190}
	(L^{s_1}(\mathbb{R})+L^{s_2}(\mathbb{R}))\cap \dot W^{1-1/s_3}_{s_3}(\mathbb{R}) \cap \dot W^{2-1/s_4}_{s_4}(\mathbb{R}),
\end{equation}
where $s_i \in \{r, r_1, r_2\}$. This form will be used during the estimate of $\nabla^2 v$.

The previous procedure of estimate the second derivatives of the velocity required introducing simplified
problem and subtracting inhomogeneity, which was connected to $\nabla p$. In our case $\nabla p\in L^{r}(\mathbb{R}) + L^{r_1}(\mathbb{R})$.
Let us denote as $w$ the solution to this simplified system, with the right hand side equal $\nabla p$. 

The first space $L^{r}(\mathbb{R})$ is more convenient for us in a sense, that since $r < 3/2$ then
Theorem \ref{OR2_80} assures that $w\in L^{r2}(\mathbb{R}^2)$, $\nabla w \in L^{r_1}(\mathbb{R}^2)$ and $\nabla^2 w \in L^r(\mathbb{R}^2)$,
which gives us that $w_{|x_2 = 0}$ is in a sum of space of the form (\ref{mainTh190}), as we have seen during previous
considerations connected with $\widetilde{d}$. This implies, that subtraction of $w$ essentially will not change
the class, where $D$ belongs to.

To deal with the part of the gradient of the pressure $\nabla p$, which belongs to $L^{r_1}$ we will have to distinguish
the case $a_2 < 0$ and $a_2 \geq 0$. Before we do this we want to notice, that since $r_1 \in (3/2, 3)$ we have:
\begin{equation}\label{mainTh200}
	\nabla^2 w\in L^{r_1}(\mathbb{R}^2) \quad \textrm{and}\quad \nabla^2 w\in L^{r_2}(\mathbb{R}^2),
\end{equation}
since $3r_1 / (3-r_1) = 3r/(3-2r) = r_2$. This assures that, independently of the signum of $a_2$, we have:
\begin{equation}\label{mainTh210}
	w_{|x_2 = 0} \in \dot W^{1-1/r_2}_{r_2}(\mathbb{R}) \cap \dot W^{2-1/r_1}_{r_1}(\mathbb{R}).
\end{equation}
In case of $a_2 < 0$ we may additionally use Theorem \ref{traceTh} to obtain, that $w \in L^{r_1}(\mathbb{R})$.

Summarizing -- the subtraction of inhomogeneity using vector field $w$ sets $\widetilde{D} = D - w$ in the following 
function spaces:
\begin{itemize}
	\item for $a_2 < 0$:
	\begin{equation}\label{mainTh220}
	\widetilde{D} \in \sum_{s_1, s_2, s_3, s_4 \in \{r, r_1, r_2\}} (L^{s_1}(\mathbb{R})+L^{s_2}(\mathbb{R}))\cap \dot W^{1-1/s_3}_{s_3}(\mathbb{R}) \cap \dot W^{2-1/s_4}_{s_4}(\mathbb{R}),
	\end{equation}
	\item for $a_2 \geq 0$:
	\begin{equation}\label{mainTh230}
	\widetilde{D} \in \sum_{s_3, s_4 \in \{r, r_1, r_2\}}\dot W^{1-1/s_3}_{s_3}(\mathbb{R}) \cap \dot W^{2-1/s_4}_{s_4}(\mathbb{R}),
	\end{equation}
\end{itemize}
with appropriate estimates.

We are now in position to obtain estimates on $\nabla^2 u$. We proceed as in the case of $\nabla p$, i.e.
we estimate particular parts of $\nabla^2 v$ by a proper part of the norm of $\widetilde{D}$. For example,
in case $a_2 < 0$ estimate of $u_{,22}$ would look like follows: 
we recall $I_1$ and $I_2$ from (\ref{lp910}). Since $I_1$ can be estimated by the $L_p$-norm of $\widetilde{D}$
and $I_2$ can be estimated by the $\dot W^{2-1/p}_p$-norm of $\widetilde{D}$, then for the part od $\widetilde{D}$, 
which belongs to, say, $(L^{r_1}+L^{r_2})\cap\dot W^{1-1/r_1}_{r_1} \cap \dot W^{2-1/r}_r$ we get an estimate
for $u_{,22}$ in the space $L^{r} + (L^{r_1} + L^{r_2})$. Similarly we may estimate other terms.
The Reader immediately notice, that in the case $a_2 = 0$ exactly the same procedure works, since
all necessary requirements on $\widetilde{D}$ are satisfied. We may thus summarize this with the following
inequality:
\begin{multline}\label{mainTh240}
	\|\nabla^2 u\|_{L^{r}(\mathbb{R}^2_+)+L^{r_1}(\mathbb{R}^2_+)+L^{r_2}(\mathbb{R}^2_+)} \leq\\
		\sum_{s_1, s_2, s_3, s_4 \in \{r, r_1, r_2\}} \|\widetilde{D}\|_{(L^{s_1}(\mathbb{R})+L^{s_2}(\mathbb{R}))\cap \dot W^{1-1/s_3}_{s_3}(\mathbb{R}) \cap \dot W^{1-1/s_4}_{s_4}(\mathbb{R})},
\end{multline}
which we shown to be valid for $a_2 \leq 0$.

For $a_2 > 0$ we encounter a small obstacle, namely during estimates we need the $\dot W^{2-2/p}_p$-norm, which
does not explicitly appear in the norm of $\widetilde{D}$. To deal with this we notice, that
the $\dot W^{2-2/p}$-norm is required in terms, which come from the multiplication in a Fourier space by a smooth
function with bounded support (see for example $I_1$ from (\ref{lp800})). Once this is known we can use 
Lemma \ref{mainTh250} to estimate the $\dot W^{2-2/s}_s$-norm with the $\dot W^{1-1/s}_s$-norm, which in our case
might be written as:
\begin{equation}\label{mainTh290}
	\|\widetilde{D}\|_{\dot W^{2-2/s_1}_{s_1}(\mathbb{R}) \cap \dot W^{2-1/s_2}_{s_2}(\mathbb{R})} \leq
		\|\widetilde{D}\|_{\dot W^{1-1/s_1}_{s_1}(\mathbb{R}) \cap \dot W^{2-1/s_2}_{s_2}(\mathbb{R})},
\end{equation}
where $s_1, s_2 \in \{r, r_1, r_2\}$. Once we have estimate of this norm we may estimate
terms in case $a_2 > 0$ in an exactly the same way it was made earlier to obtain, that
(\ref{mainTh240}) is valid also for $a_2 > 0$.

Summarizing, we have proved the following inequality:
\begin{multline}\label{mainTh300}
	\|\nabla^2 u\|_{L^{r}(\mathbb{R}^2_+)+L^{r_1}(\mathbb{R}^2_+)+L^{r_2}(\mathbb{R}^2_+)} \leq\\
		C\sum_{s_1, s_2, s_3, s_4 \in \{r, r_1, r_2\}} \|\widetilde{D}\|_{(L^{s_1}(\mathbb{R})+L^{s_2}(\mathbb{R}))\cap \dot W^{1-1/s_3}_{s_3}(\mathbb{R}) \cap \dot W^{1-1/s_4}_{s_4}(\mathbb{R})},
\end{multline}
which, together with previous estimates, gives us the following inequality for the solution $(v, q)$ to the system (\ref{lpa0})-(\ref{lpa30}):
\begin{multline}\label{mainTh310}
	\|\nabla q\|_{L^r(\mathbb{R}^2_+) + L^{r_1}(\mathbb{R}^2_+)} +
		\|\nabla^2 v\|_{L^r(\mathbb{R}^2_+) + L^{r_1}(\mathbb{R}^2_+) +L^{r_2}(\mathbb{R}^2_+)} \leq \\
			C\left(\|F\|_{L^r(\mathbb{R}^2_+)} + \|G\|_{W^{1}_r(\mathbb{R}^2_+)} +
					\|\underline{b}\|_{W^{1-1/r}_r(\mathbb{R})} + \|\underline{d}\|_{W^{2-1/r}_r(\mathbb{R})}\right).
\end{multline}
We now recall the fact, that $r \in (1, 3/2)$, which implies that $r_1 = 3r/(3-r) > 3/2$ and $r_2 = 3r/(3-2r) > 3$.
We also know, that the support of $q$ and $v$ is compact, since this came from the localization procedure, hence
$L^{r_1}$ and $L^{r_2}$ norm majorize $L^r$ norm, with a coefficient dependent only on the size of the support of
$q$ and $v$, thus the following inequality holds:
\begin{multline}\label{mainTh320}
	\|\nabla q\|_{L^r(\mathbb{R}^2_+)} +
		\|\nabla^2 v\|_{L^r(\mathbb{R}^2_+)} \leq \\
			C\left(\|F\|_{L^r(\mathbb{R}^2_+)} + \|G\|_{W^{1}_r(\mathbb{R}^2_+)} +
					\|\underline{b}\|_{W^{1-1/r}_r(\mathbb{R})} + \|\underline{d}\|_{W^{2-1/r}_r(\mathbb{R})}\right).
\end{multline}
This estimate allows us to complete the proof of Theorem \ref{mainTh}, since, as we have shown earlier,
this proof requires estimates in the whole space, which is guaranteed due to Theorem \ref{OR2_40},
and local estimates near the boundary, which we have just proved.
Thus, the proof of Theorem \ref{mainTh} is completed.

\section{Appendix}


In this section we give statements of lemmas and theorems, which were used in proofs of the previous results.
The following two Theorems are extensively used in our paper. The first one is due to Marcinkiewicz:
\begin{Theorem}\label{Marcinkiewicz}
  Suppose that the function $\Phi : \mathbb{R}^m \to \mathbb{C}$ is smooth enough and there exists
  $M > 0$ such that for every point $x\in \mathbb{R}^m$ we have
  \begin{equation}
  	|x_{j_1}\ldots x_{j_k}| \left|\frac{\partial^k\Phi}{\partial x_{j_1}\ldots \partial x_{j_k}}\right|\leq M,
  	\quad 0\leq k\leq m, 1\leq j_1 < \ldots < j_k \leq m.
  \end{equation}
  Then the operator
  \begin{equation}
  	Pg(x) = (2\pi)^{-m}\int_{\mathbb{R}^m} dy e^{ix\cdot y}\Phi(y)\int_{\mathbb{R}^m}e^{-iy\cdot z}g(z)dz
  \end{equation}
  is bounded in $L_p(\mathbb{R}^m)$ and
  \begin{equation}
  	\|Pg\|_{L_p(\mathbb{R}^m)} \leq A_{p, m} M \|g\|_{L_p(\mathbb{R}^m)}
  \end{equation}
\end{Theorem}
The next theorem is due to Lizorkin:
\begin{Theorem}\label{Lizorkin}
  Let
  \begin{equation}\label{multiTransform}
    Tu \equiv h(x) = \frac{1}{2\pi}\int_{R^2}e^{ix\cdot\xi}\Phi(\xi)\hat{u}(\xi)d\xi,
  \end{equation}
  where $\Phi:\mathbb{R}^2\to\mathbb{R}^2$ is continous together with the derivatives
  $$\frac{\partial\Phi}{\partial\xi_1}, \frac{\partial\Phi}{\partial\xi_2}, \frac{\partial^2\Phi}{\partial\xi_1\partial\xi_2},$$
  for $|\xi_i|>0$, $i = 1, 2$. Then, if for some $\beta\in[0, 1)$ and $M>0$
  \begin{equation}\label{OR2_20}
    |\xi_1|^{\kappa_1+\beta}|\xi_2|^{\kappa_2+\beta}
       \left|\frac{\partial^{\kappa}}{\partial\xi_1^{\kappa_1}\partial\xi_2^{\kappa_2}}\right| \leq M,
  \end{equation}
  where $\kappa_i$ is zero or one and $\kappa = \kappa_1+\kappa_2$, the integral transform (\ref{multiTransform})
  defines a bounded linear operator from $L^q(\mathbb{R}^2)$ into $L^r(\mathbb{R}^2)$, $1 < q < \infty$,
  $1/r = 1/q - \beta$, and we have:
  \begin{equation}\label{OR2_30}
    \|Tu\|_{L^r}\leq C\|u\|_{L^q},
  \end{equation}
  with a constant $C = c(q, r)M$.
\end{Theorem}

The following Lemma allows us to estimate a homogeneous norm of a function on a boundary:
\begin{Lemma}{}\label{W1-1p}
	Let $f\in L^{s}(\mathbb{R}^2)$ and $\nabla f\in L^m(\mathbb{R}^2)$. For $s \in (1, 2]$ we assume $m\in (1, s)$,
	and for $s > 2$ we assume $m\in (\frac{2s}{2+s}, s)$. 
	Then $f_{|x_2 = 0} \in \dot W^{1-1/m}_m(\mathbb{R})$ and the following inequality holds:
	\begin{equation}\label{}
		\|f\|_{\dot W^{1-1/m}_m(\mathbb{R}} \leq C \|\nabla f\|_{L^m(\mathbb{R}^2)}.
	\end{equation}
\end{Lemma}
\begin{Proof}{}
	We construct a sequence of functions, which converge to $f$ appropriately and their trace is in a proper function space.
	Let us introduce a smooth cut-off function $\eta(x)$ such that:
	$\eta(x) = 1$ for all $x\in B(0, 1)$ and $\eta(x) = 0$ for all $x\in \mathbb{R}^2\setminus B(0, 2)$, together with
	sequence of cut-off functions $\eta_k(x)$, defined as $\eta_k(x) = \eta(x/k)$.
	
	Let $f_k(x) = f(x)\eta_k(x)$. Since $\eta_k(x)$ has a bounded support and $m < 2$ we have $f_k(x) \in W^1_m(\mathbb{R}^2)$
	and hence $f_k(x)_{|x_2 = 0} \in W^{1-1/m}_m(\mathbb{R})$. Moreover, from the standard scaling argument it is easy to see, that
	$\|f_k(x)_{|x_2 = 0}\|_{\dot W^{1-1/m}_m(\mathbb{R})} \leq \|\nabla f_k(x)\|_{L^m(\mathbb{R}^2)}$.
	
	Of course $f_k \to f$ in $L^{s}(\mathbb{R}^2) $ as $k\to\infty$. To prove our theorem we need to show that ${f_k}_{|x_2 = 0}$
	is a Cauchy sequence in $\dot W^{1-1/m}_m(\mathbb{R})$. From the definition of $f_k$ we get:
	\begin{eqnarray}
		\|f_k-f_l\|_{\dot W^{1-1/m}_m(\mathbb{R})} & \leq & \|\nabla f_k - \nabla f_l\|_{L^m(\mathbb{R}^2)} \\
			& \leq & \|\nabla f (\eta_k - \eta_l)\|_{L^m(\mathbb{R}^2)} + \|f \nabla (\eta_k - \eta_l)\|_{L^m(\mathbb{R}^2)}.
	\end{eqnarray}
	The first term on the right hand side is obviously small for large $k$ and $l$. The second is also small for $k$ and $l$
	large enough, since
	\begin{equation}
		\|f \nabla (\eta_k - \eta_l)\|_{L^m(\mathbb{R}^2)} \leq \|u\|_{L^s(\mathbb{R}^2)} \|\nabla(\eta_k - \eta_l)\|_{L^{mn/(n-m)}(\mathbb{R}^2)},
	\end{equation}
	and $\|\nabla\eta_k\|_{L^{mn/(n-m)}(\mathbb{R}^2)}\to 0$ as $k\to\infty$. Indeed, $|\textrm{supp} \nabla \eta_k| \sim k^2$ and
	$|\nabla \eta_k| \sim 1/k$, hence $\|\nabla\eta_k\|_{L^{mn/(n-m)}(\mathbb{R}^2)} \leq C(\eta) k^{(2-mn/(n-m))/r} \to 0$ as $k\to\infty$,
	since under our assumptions $2-mn/(n-m) < 0$.
\end{Proof}

We use the following lemma to set a function space, where the trace of a function belongs to:
\begin{Lemma}\label{mainTh100}
	Let $f \in L^{p_1}(\mathbb{R}^2_+)$ and $\nabla f \in L^{p_2}(\mathbb{R}^2_+)$, then
	$f_{|x_2 = 0} \in L^{p_1}(\mathbb{R}) + L^{p_2}(\mathbb{R})$ and the following estimate is valid:
	\begin{equation}
		\|f_{|x_2 = 0}\|_{L^{p_1}(\mathbb{R}) + L^{p_2}(\mathbb{R})} \leq C(\|f\|_{L^{p_1}(\mathbb{R}^2_+)} + \|\nabla f\|_{L^{p_1}(\mathbb{R}^2_+)}).
	\end{equation}
\end{Lemma}
\begin{Proof}{}
	Introducing a smooth cut-off function $\eta(x_2)$ such that $\eta(x_2) = 1$ for $x_2 < 1$ and $\eta(x_2) = 0$ for $x_2 > 2$
	we can write:
	\begin{multline}
		f(0, x') = \eta(0)f(0, x') = \int_0^2 \partial_{x_2} (\eta(s)f(s, x'))ds =\\
			=\int_0^2 \eta'(s) f(s, x') ds + \int_0^2 \eta \partial_{x_2} f(s, x') ds.
	\end{multline}
	This proves, that $f(0, x')$ is a sum of two functions from $L^{p_1}(\mathbb{R}^2_+)$ and $L^{p_2}(\mathbb{R}^2_+)$, which
	completes the proof of the Lemma.
\end{Proof}

The following Lemma is substantial to estimate higher homogeneous norms of function with bounded support in Fourier space:
\begin{Lemma}\label{mainTh250}
	Let $f \in \dot W^s_r(\mathbb{R})$, $s\notin \mathbb{Z}$. Given a smooth function $\pi(k)$ such that
	$\pi(k) = 1$ for $|k| \leq L$ and $\pi(k) = 0$ for $|k| \geq L+1$. Then
	$\mathcal{F}^{-1}_{k}\left(\pi(k)\hat f\right) \in \dot W^{s+\epsilon}_r(\mathbb{R})$ and
	the following inequality holds:
	\begin{equation}\label{mainTh260}
		\|\mathcal{F}^{-1}_{k}\left(\pi(k)\hat f\right)\|_{\dot W^{s+\epsilon}_r(\mathbb{R})} \leq
			C(\epsilon) \|f\|_{\dot W^{s}_r(\mathbb{R})},
	\end{equation}
	where $\epsilon > 0$ is an arbitrary positive constant.
\end{Lemma}

\begin{Proof}{}
	In case of $s\notin \mathbb{Z}$ we have $\dot W^{s}_r(\mathbb{R}) = \dot B^s_{r, r}(\mathbb{R})$,
	where $\dot B^s_{r, r}(\mathbb{R})$ stands for the homogeneous Besov space equipped
	with a norm:
	\begin{equation}\label{mainTh270}
		\|f\|_{\dot B^s_{r, r}(\mathbb{R})} = 
	\left(\sum_{j = -\infty}^\infty 2^{jsr} \left\|\mathcal{F}^{-1}_{k}\left(\varphi_j \hat f\right)\right\|_{L^r(\mathbb{R})}^r\right)^{1/r},
	\end{equation}
	where $\{\varphi_j\}_{j = -\infty}^\infty$ is a set of smooth function, each of them of bounded support
	$\mathrm{supp~}\varphi_j \subset \{\xi : 2^{j-1} \leq |\xi|\leq 2^{j+1}\}$ and such that
	$\sum_{j = -\infty}^\infty \varphi_j(\xi) = 1$ for every $\xi\in \mathbb{R}\setminus \{0\}$ (see \cite{Triebel}).
	
	Multiplication by the function $\pi$ implies that the sum in (\ref{mainTh270}), corresponding
	to the function $\mathcal{F}^{-1}_{k}\left(\pi\hat f\right)$, has infinite number of elements
	with nonpositive $j$, and finite number of elements of elements with positive $j$. 
	Without loss in generality we can assume, that $L > 1$. Then, in the case of negative $j$ we have:
	\begin{multline}\label{mainTh275}
		\sum_{j = -\infty}^0 2^{j(s+\epsilon)r} \|\mathcal{F}^{-1}_{k}\left(\varphi_j \pi \hat f\right)\|_{L^r(\mathbb{R})}^r =
			\sum_{j = -\infty}^0 2^{j\epsilon r}2^{jsr}\|\mathcal{F}^{-1}_{k}\left(\varphi_j \pi \hat f\right)\|_{L^r(\mathbb{R})}^r \leq\\
			\leq \sum_{j = -\infty}^0 2^{jsr}\|\mathcal{F}^{-1}_{k}\left(\varphi_j \pi \hat f\right)\|_{L^r(\mathbb{R})}^r
			= \sum_{j = -\infty}^0 2^{jsr}\|\mathcal{F}^{-1}_{k}\left(\varphi_j \hat f\right)\|_{L^r(\mathbb{R})}^r \leq \|f\|_{\dot B^{s}_{r, r}}^r,
	\end{multline}
	since $\pi(\xi) = 1$ for $\xi \in \cup_{j = -\infty}^0 \textrm{supp~}\varphi_j$.
	Remaining terms (finite number) can be estimated using Marcinkiewicz theorem:
	\begin{multline}\label{mainTh277}
		\sum_{j = 1}^{\lceil 1+\log_2(L+1)\rceil}
		2^{j(s+\epsilon)r}\|\mathcal{F}^{-1}_{k}\left(\varphi_j \pi \hat f\right)\|_{L^r(\mathbb{R})}^r \leq\\
			\sum_{j = 1}^{\lceil 1+\log_2(L+1)\rceil}2^{j\epsilon r} C(\pi) 2^{jsr}\|\mathcal{F}^{-1}_{k}\left(\varphi_j \hat f\right)\|_{L^r(\mathbb{R})}^r \leq
				(2L+2)^{\epsilon r}C(\pi)\|f\|_{\dot B^s_{r,r}(\mathbb{R})}^r.
	\end{multline}
	This completes the proof of the following inequality:
	\begin{equation}\label{mainTh280}
		\left\|\mathcal{F}^{-1}_{k}\left(\pi \hat f\right)\right\|_{\dot W^{s+\epsilon}_{r}(\mathbb{R})}
			\leq C(\pi)(2L+3)^{\epsilon}\|f\|_{\dot W^{s}_{r}(\mathbb{R})},
	\end{equation}
	and the proof of Lemma \ref{mainTh250}.
\end{Proof}

\begin{footnotesize}
{\it \bf Acknowledgement. }
The author would like to thank Piotr Mucha for useful discussions during preparation of this paper and for his great patience.
The paper has been supported by Polish grant No. N201 035 32/2271.	
\end{footnotesize}


\begin{thebibliography}{99}


\bibitem[1]{ADN1} Agmon, S., Douglis, A., Nirenberg, L., 
Estimates near the boundary for solutions of elliptic partial differential equations satisfying general boundary conditions. I.,
Comm. Pure Appl. Math. 12 (1959) 623--727.

\bibitem[2]{ADN2} Agmon, S., Douglis, A., Nirenberg, L., 
Estimates near the boundary for solutions of elliptic partial differential equations satisfying general boundary conditions. II.,
Comm. Pure Appl. Math.  17 (1964) 35--92. 

\bibitem[3]{Amick1988} Amick, C.J.: On Leray's Problem of Steady Navier-Stokes
Flow Past a Body in the Plane, Acta Math., 161 (1988), 71--130.

\bibitem[4]{BorPil} Borchers, W., Pileckas, K., Note on the Flux Problem for Stationary
Incompressible Navier-Stokes Equations in Domains with Multiply Connected
Boundary, Acta Appl. Math. 37 (1994), 21--30.

\bibitem[5]{Farwig} Farwig, R.: Stationary solutions of compressible Navier-Stokes
equations with slip boundary conditions, Comm. PDE 14, (1989) 1579--1606

\bibitem[6]{FiSm} Finn, R., Smith, D.R.; On the Stationary Solution of the
Navier-Stokes Equations in Two Dimensions, Arch. Rational Mech. Anal. 25 (1967)
26--39.

\bibitem[7]{Fujita} Fujita, H., Remarks on the Stokes flow under slip and leak boundary 
conditions of friction type.  Topics in mathematical fluid mechanics,  73--94,
 Quad. Mat., 10,  2002.
 
\bibitem[8]{Galdi1993} Galdi, G.P.: Existence and Uniqueness at Low Reynolds
Number of Stationary Plane Flow of a Viscous Fluid in Exterior Domains.
Recent Developments in Theoretical Fluid Mechanics, Galdi, G.P., and
Necas, J., Eds., Pitman Research Notes in Mathematics Series,
Longman Scientific and Technical, Vol. 291 (1993), 1--33.

\bibitem[9]{Galdi} Galdi, G.P.: An Introduction to the Mathematical Theory of the
Navier-Stokes Equations, Springer Tracts in Natural Philosophy, 1994.

\bibitem[10]{GS} Galdi, G.P.; Sohr, H.: On the asymptotic structure 
of plane steady flow of a viscous fluid in exterior domains.  
Arch. Rational Mech. Anal.  131  (1995),  no. 2, 101--119.

\bibitem[11]{GW1} Gilbarg, D.; Weiberger, H.F.: Asymptotic Properties of Leray's
Solution of the Stationary Two-Dimensional Navier-Stokes Equations.
Russian Math. Surveys, 29 (1974), 109--123.

\bibitem[12]{GW2} Gilbarg, D.; Weiberger, H.F.: Asymptotic Properties of Steady Plane
Solutions of the Navier-Stokes Equations with Bounded Dirichlet Integral.
Ann. Scuola Norm. Sup. Pisa, (4), 5 (1978), 381--404.

\bibitem[13]{Hopf} Hopf, E., Ein allgemeiner Endlichkeitssatz der Hydrodynamik, 
Math. Ann. 117 (1941), 764--775.

\bibitem[14]{Itoh} Itoh, S.; Tanaka N.; Tani A.: The initial value problem
for the Navier-Stokes equations with general slip boundary
condition, Adv. Math. Sci. Appl. 4, (1994) 51--69

\bibitem[15]{2dextlin} Konieczny, P., Linear flow problems in 2D exterior
domain for 2D incompressible fluid flows, Banach Center Publ., to appear.

\bibitem[16]{pkpbm} Konieczny, P.; Mucha, P. B., On nonhomogeneous slip boundary conditions
for 2D incompressible fluid flows, Internat. J. Engrg. Sci. 44 (2006), no. 11-12, 738--747.

\bibitem[17]{Konieczny} Konieczny, P., On a steady flow in a three dimensional infinite pipe,
Coll. Math. 104 (2006), no. 1, 33--56.
	
\bibitem[18]{Ladyz} Ladyzhenskaya, O.A.: The Mathematical Theory of Viscous
Incompressible Flow, Gordon and Breach, New York, 1966
			 		  
\bibitem[19]{Mucha1} Mucha, P.B., On the inviscid limit of the Navier-Stokes 
equations for flows with large flux, Nonlinearity 16 (2003), 1715--1732. 		

\bibitem[20]{Mucha2} Mucha, P.B., Asymptotic behavior of a steady flow 
in a two-dimensional pipe. Studia Math. 158 (2003), no. 1, 39--58.
					  
\bibitem[21]{MuRa} Mucha, P. B.; Rautmann, R., Convergence of Rothe's scheme for 
the Navier-Stokes equations wish slip boundary conditions in 2D domains.
Z. Angew. Math. Mech. 86 (2006), no. 9, 691--701.

\bibitem[22]{ZaMuJDE} Mucha, P. B.; Zaj\c{a}czkowski, W. M., 
On a $L\sb p$-estimate for the linearized compressible Navier-Stokes equations with the Dirichlet boundary conditions.  
J. Differential Equations 186 (2002),  no. 2, 377--393.

\bibitem[23]{ZaMuStud} Mucha, P. B.; Zaj\c{a}czkowski, W. M., 
On the existence for the Cauchy-Neumann problem for the Stokes system in the $L\sb p$-framework.  
Studia Math.  143  (2000),  no. 1, 75--101.

\bibitem[24]{PokornyPHD} Pokorn\'y, M., Asymptotic behaviour of solutions to certain PDE's
describing the flow of fluids in unbounded domains. Ph.D. thesis, Charles University, Prague
\& University of Toulon and Var, Toulon-La Garde, 1999.

\bibitem[25]{Solonnikov} Solonnikov, V. A. 
Estimates for solutions of a non-stationary linearized system of Navier-Stokes equations. (Russian)  
Trudy Mat. Inst. Steklov.  70  (1964) 213--317.

\bibitem[26]{Triebel} Triebel, H.: Theory of function spaces,
Mathematik und ihre Anwendungen in Physik und Technik, 38. 
Akademische Verlagsgesellschaft Geest \& Portig K.-G., Leipzig, 1983.

\end{thebibliography}
\end{document}